\documentclass[aps,preprintnumbers,eqsecnum,prb,preprint,11pt]{revtex4}
\usepackage[T1]{fontenc}
\usepackage{amsfonts,amsmath,amsbsy,amssymb}
\usepackage[dvips]{graphicx}

\input{tcilatex}

\begin{document}
\baselineskip=14pt
\title{Intrinsic Zeeman Effect in Graphene}
\author{Motohiko Ezawa}

\affiliation{Department of Physics, University of Tokyo, Hongo 7-3-1, 113-0033, Japan }
\date{April 14, 2007}

\begin{abstract}
The intrinsic Zeeman energy is precisely one half of the
cyclotron energy for electrons in graphene.
As a result a Landau-level mixing occurs 
to create the energy spectrum comprised of the $4j$-fold degenerated
zero-energy level and $4$-fold degenerated nonzero-energy levels 
in the $j$-layer graphene,
where $j=1,2,3$ for monolayer, bilayer and trilayer, respectively. 
The degeneracy manifests itself in the quantum Hall (QH) effect.
We study how the degeneracy is removed by the Coulomb interactions.
With respect to the zero-energy level, 
an excitonic gap opens by making a BCS-type condensation of electron-hole pairs 
at the filling factor $\nu =0$. 
It gives birth to the Ising QH ferromagnet at $\nu =\pm 1$ for monolayer, $\nu
=\pm 1,\pm 3$ for bilayer, and $\nu =\pm 1,\pm 3,\pm 5$ for trilayer graphene from the zero-energy degeneracy. 
With respect to the nonzero-energy level,
a remarkable consequence is derived that the effective Coulomb potential
depends on spins, since a single energy level contains up-spin and
down-spin electrons belonging to different Landau levels.  
The spin-dependent Coulomb interaction leads to the valley polarization at $\nu =\pm 4, \pm 8, \pm 12, \cdots$ for monolayer, 
$\nu =\pm 2, \pm 6, \pm 10, \cdots$ for bilayer, 
and $\nu =\pm 2,\pm 4, \pm 8, \pm 12, \cdots$ for trilayer graphene.
\end{abstract}
\maketitle

\section{Introduction}

\begin{figure}[t]
\begin{center}
\includegraphics[width=0.68\textwidth]{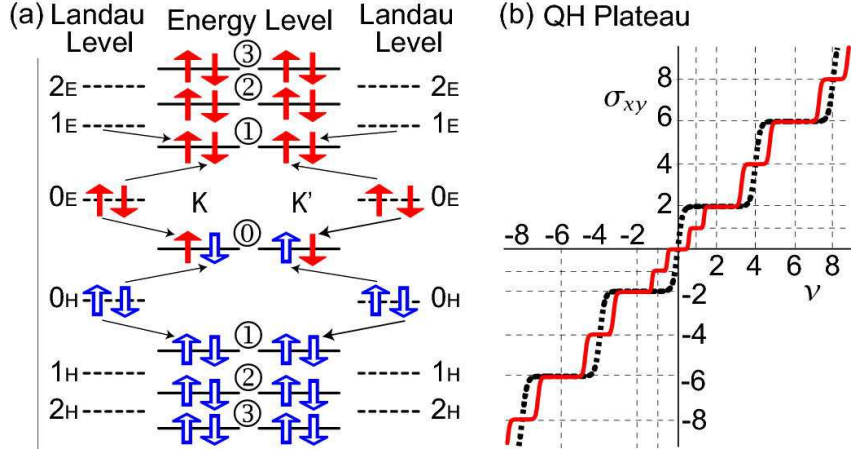}
\end{center}
\caption{\baselineskip=12pt{}(Color online) (a) The energy level (number in circle) and the
Landau level (number with index E or H) for electrons and holes. The spin is
indicated by a solid red (open blue) arrow for electron and hole at the K
and K' points. The $N$th energy level contains up(down)-spin electrons from
the $N$th Landau level and down(up)-spin electrons from the ($N$-$1$)th
Landau level at the K (K') point for $N\geq 1$. The zero-energy level ($N=0$%
) contains up(down)-spin electrons and down(up)-spin holes at the K (K')
point. (b) The QH conductivity in graphene. The dotted black curve shows the
sequence $\protect\nu =\pm 2,\pm 6,\pm 10,\cdots $, while the solid red
curve the sequence $\protect\nu =0,\pm 1,\pm 2,\pm 4,\cdots $. }
\label{FigGraphLevel}
\end{figure}

Recent experiments has established that the charge carriers in graphene are
massless Dirac electrons\cite{Nov1,Nov2,Zhang,Nov3,Zhang06L}. Dirac
electrons have a peculiar property that they have the intrinsic Zeeman
energy precisely one half of the cyclotron energy in magnetic field.
Consequently a Landau level mixing occurs so that one energy level contains
up-spin and down-spin electrons coming from different Landau levels. It has
two important consequences; the emergence of the zero-energy state\cite%
{McClure}, and the degeneracy of the up-spin and down-spin states for each
nonzero-energy level [Fig.\ref{FigGraphLevel}(a)]. The aim of this paper is
to explore new phenomena due to this intrinsic Zeeman effect in graphene.

The quantum Hall effect (QHE) in graphene\cite{Nov1,Nov2,Zhang,Nov3,Zhang06L}
is unconventional. The filling factors form a series [Fig.\ref{FigGraphLevel}%
(b)], 
\begin{equation}
\nu =0,\pm 1,\boldsymbol{\pm 2},\pm 4,\boldsymbol{\pm 6},\pm 8,\boldsymbol{\pm 10},\pm
12,\cdots ,  \label{G-SerieMono}
\end{equation}%
where the bold-face series had been predicted\cite%
{Ando02B,Gusynin05L,Peres06B} before it was found experimentally\cite%
{Nov1,Nov2,Zhang,Nov3}, while the full series was discovered later\cite%
{Zhang06L} with larger magnetic field applied.\ Subsequently theoretical
works\cite{Fisher,Nomura,Goerbig,Gusynin06B,Ezawa} have been made to
interpret the series. Furthermore, the series reads%
\begin{equation}
\nu =0,\pm 1,\pm 2,\pm 3,\boldsymbol{\pm 4},\pm 6,\boldsymbol{\pm 8},\pm 10,\boldsymbol{%
\pm 12},\cdots ,  \label{G-SerieBL}
\end{equation}%
in a bilayer graphene, where the bold-face series has been found
experimentally\cite{Nov3} and studied theoretically\cite{McCann}, while it
reads%
\begin{equation}
\nu =0,\pm 1,\pm 2,\pm 3,\pm 4,\pm 5,\boldsymbol{\pm 6},\pm 8,\boldsymbol{\pm 10}%
,\pm 12,\cdots ,  \label{G-SerieTL}
\end{equation}%
in a trilayer graphene, where the bold-face series has been predicted
theoretically\cite{Guinea}. In this paper we show that the full series
emerge when Coulomb interactions become important. It is notable that the
basic height in the Hall conductance is $4e^{2}/h$ for the bold-face series
for all these graphene systems, indicating the 4-fold degeneracy of the
energy level except for the first step at the $\nu =0$ point within
noninteracting theory.

\begin{figure}[t]
\begin{center}
\includegraphics[width=0.7\textwidth]{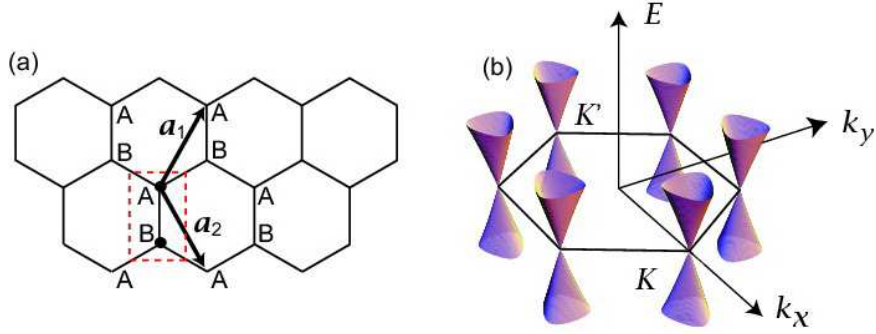}
\end{center}
\caption{\baselineskip=12pt{}(Color online) (a) Graphene is made of a honeycomb lattice. It
consists of two triangular sublattices generated by the two basis vectors $%
\boldsymbol{a}_{1}$ and $\boldsymbol{a}_{2}$ from the base points A and B in
primitive cell (dotted rectangle). (b) The reciplocal lattice is also a
honeycomb lattice. The first Brillouin zone is depicted together with the
Dirac valleys describing the low-energy band structure.}
\label{FigABcone}
\end{figure}

Conduction and valence bands in graphene form conically shaped valleys,
touching at a point [Fig.\ref{FigABcone}]. There are two inequivalent
Brillouin zone corners, called the K and K' points, at which massless Dirac
electrons emerge\cite{Slonczewski,Semenoff,Ajiki}. Let us refer to the
valley as the Dirac valley, and assign the valley index to the electron so
that the electron at the K (K') point carries the index $\tau =+$ ($-$). As
we have mentioned, the graphene QH system is characterized by the emergence
of the zero-energy state and the degeneracy of the up-spin and down-spin
states for each nonzero-energy level. Since this holds separately at the K
and K' points, each energy level has a 4-fold degeneracy [Fig.\ref%
{FigGraphLevel}(a)], and the noninteracting system has the SU(4) symmetry.

Intriguing phenomena occur when we introduce Coulomb interactions. It is
necessary to consider the Coulomb effect in the zero-energy state and
nonzero-energy states separately.

The zero-energy state is distinctive, since it contains both electrons and
holes. Electron-hole pairs form an excitonic condensation\ due to attractive
interaction, producing an excitonic gap. We obtain a BCS-type state of
electron-hole pairs at $\nu =0$. As a consequence of excitonic condensation,
the degeneracy of the zero-energy state is resolved into two subbands each
of which contains either electrons or holes. The Coulomb Hamiltonian,
projected to each of them, has the U(1) symmetry but is broken into the Z$%
_{2}$ symmetry. We obtain the Ising QH ferromagnet at $\nu =\pm 1$.

We next study nonzero-energy states of electrons, where a single energy
level contains up-spin and down-spin electrons belonging to different Landau
levels [Fig.\ref{FigGraphLevel}(a)]. We derive a remarkable consequence that
the effective Coulomb potential depends on the spin and the valley in each
energy level. Namely, it depends on the spin of electrons as well as on
which valley electrons belongs to. The Coulomb Hamiltonian, projected to a
single energy level, possesses only the U(1)$\otimes $U(1)$\otimes $Z$_{2}$
symmetry. Consequently, the Coulomb interaction resolves the 4-fold
degeneracy into two 2-fold degeneracies, generating a new series at $\nu
=\pm 4,\pm 8,\pm 12,\cdots $. The ground state is a valley polarized state,
by which we mean that more electrons are present in one valley than the
other. This occurs because the Coulomb energy is lower in higher Landau
levels.

Our analysis can be generalized to the $j$-layer graphene system, where $j=1$%
, $2$ and $3$ correspond to monolayer, bilayer and trilayer, respectively.
We show that the zero-energy states have the $4j$-fold degeneracy. As a
result the quantized values of the Hall conductivity become%
\begin{equation}
\sigma _{xy}=\pm \left( n+\frac{j}{2}\right) \frac{4e^{2}}{h},\qquad
n=0,1,2,\cdots ,  \label{HallConduA}
\end{equation}%
within noninteracting theory. We also discuss the effects due to Coulomb
interactions.

This paper is composed as follows. In Section \ref{SecSUSY} we investigate
the energy spectrum in the presence of magnetic field. We show that two
Landau levels mix to create one energy level. In Section \ref{SecGraphCoulo}
we construct the projected Coulomb Hamiltonian relevant to analyze physics
in the $N$th energy level. The effective Coulomb potential is shown to
depend on the spin and the valley through the form factors characterizing
Landau levels. In Section \ref{SecExciton} we treat the zero-energy level ($%
N=0$), which contains both electrons and holes. We present a clear-cut
description of an excitonic condensation, producing a gap to the electron
and hole states, on the analogy of the BCS superconductor. In Section \ref%
{SecKKasymm}, studying how Coulomb interactions resolve the degeneracy of
the nonzero-energy level, we derive the valley polarized ground state. In
Section \ref{SecMultiGraph} our analysis is generalized to multilayer
graphene systems. Finally, Section \ref{SecDiscu} is devoted to discussions.

\section{Dirac Electrons and Intrinsic Zeeman Effect}

\label{SecSUSY}

\subsection{Dirac Hamiltonian}

The low-energy band structure of graphene is described by conically shaped
valleys, touching at a point. There exists two inequivalent Brillouin zone
corners $\pm \boldsymbol{K}=(\pm 4\pi /\sqrt{3}a,0)$ with $a$ the lattice
constant. They are called the K and K' points [Fig.\ref{FigABcone}(b)]. The
two-dimensional energy dispersion relation is linear in these Dirac valleys,%
\begin{equation}
\mathcal{E}_{\tau }\left( \boldsymbol{k}\right) =\hbar v_{\text{F}}|\boldsymbol{k}%
-\tau \boldsymbol{K}|,
\end{equation}%
where $v_{\text{F}}$ is the Fermi velocity, and $\tau =\pm $ is the valley
index. This linear behavior has been confirmed experimentally\cite{Zhou} up
to 3eV. It is convenient to make the change of variable, $\boldsymbol{k}=\boldsymbol{%
k}^{\prime }+\tau \boldsymbol{K}$, and rewrite the dispersion relation as 
\begin{equation}
\mathcal{E}(\boldsymbol{k}^{\prime })=\hbar v_{\text{F}}|\boldsymbol{k}^{\prime }|%
\boldsymbol{.}  \label{DispeDirac}
\end{equation}%
We use the variable $\boldsymbol{k}^{\prime }$ to show that $\hbar \boldsymbol{k}%
^{\prime }$ is the relative momentum of electrons, $|\boldsymbol{k}^{\prime
}|\ll |\boldsymbol{K}|$, measured from $\tau \boldsymbol{K}$ at the K or K' point.

There exists one electron per one carbon and the band-filling factor is 1/2
in graphene. Hence we have electron excitations in the conduction band and
hole excitations in the valence band. Furthermore there exists the
electron-hole symmetry.

The dispersion relation (\ref{DispeDirac}) is that of `relativistic' Dirac
fermions\cite{Slonczewski,Semenoff,Ajiki}. Hence the second-quantized
Hamiltonian is given by

\begin{equation}
H=\sum_{\tau =\pm }\int \!d^{2}x\,\Psi _{\tau }^{\dagger }(\boldsymbol{x})H_{%
\text{D}}^{\tau }\Psi _{\tau }(\boldsymbol{x}),  \label{GraphHamilD}
\end{equation}%
where $H_{\text{D}}^{\tau }$ is the quantum mechanical Hamiltonian,%
\begin{equation}
H_{\text{D}}^{\tau }=v_{\text{F}}\left( \tau \sigma _{x}p_{x}+\sigma
_{y}p_{y}\right) \gamma _{5},  \label{DiracHamilD}
\end{equation}%
with $p_{i}\equiv -i\hbar \partial _{i}$, the Pauli matrix $\sigma _{i}$ and
the Dirac matrix $\gamma _{5}$. The field $\Psi _{\tau }(\boldsymbol{x})$
consists of eight components, corresponding to the spin degree of freedom,
the electron-hole degree of freedom and the valley degree of freedom. Being
massless, the Hamiltonian describes Weyl fermions, and the spin index $%
\sigma $ represents the helicity.

The field operator $\Psi _{\tau }(\boldsymbol{x})$ is expanded in terms of the
eigenfunctions of $H_{\text{D}}^{\tau }$ as 
\begin{equation}
\Psi _{\tau }(\boldsymbol{x})=\Psi _{\text{e}\tau }(\boldsymbol{x})+\Psi _{\text{h}%
\tau }(\boldsymbol{x}),
\end{equation}%
with
\begin{subequations}
\begin{align}
\Psi _{\text{e}\tau }(\boldsymbol{x})& =\sum_{\sigma }\int {\frac{d^{2}k^{\prime
}}{2\pi }}c_{\tau }^{\sigma }(\boldsymbol{k}^{\prime })u_{\sigma }^{\tau }(%
\boldsymbol{k}^{\prime })e^{i\boldsymbol{k}^{\prime }\boldsymbol{x}},
\label{GraphFieldA} \\
\Psi _{\text{h}\tau }(\boldsymbol{x})& =\sum_{\sigma }\int {\frac{d^{2}k^{\prime
}}{2\pi }}d_{\tau }^{\sigma \dag }(\boldsymbol{k}^{\prime })v_{\sigma }^{\tau }(%
\boldsymbol{k}^{\prime })e^{-i\boldsymbol{k}^{\prime }\boldsymbol{x}},
\end{align}%
where $c_{\tau }^{\sigma }(\boldsymbol{k}^{\prime })$ and $d_{\tau }^{\sigma
\dag }(\boldsymbol{k}^{\prime })$ are the annihilation operator of an electron
and the creation operator of a hole, respectively, while $u_{\sigma }^{\tau
}(\boldsymbol{k}^{\prime })$ and $v_{\sigma }^{\tau }(\boldsymbol{k}^{\prime })$ are
the corresponding eigenfunctions of the Dirac operator (\ref{DiracHamilD})
with the eigenvalue $\mathcal{E}(\boldsymbol{k}^{\prime })=\pm \hbar v_{\text{F}%
}|\boldsymbol{k}^{\prime }|$.

The field $\Psi _{\tau }(\boldsymbol{x})$ describes solely the dynamics of
electrons localized in each Dirac valley at the K or K' point. In the
dispersion relation (\ref{DispeDirac}), $\hbar \boldsymbol{k}^{\prime }$ is the
momentum fluctuation around $\pm \hbar \boldsymbol{K}$ at the K or K' point. The
total momentum of the electron is not $\hbar \boldsymbol{k}^{\prime }$ but $%
\hbar \boldsymbol{k}=\hbar \boldsymbol{k}^{\prime }\pm \hbar \boldsymbol{K}$. Due to
this fact, the eigenfunctions $u_{\sigma }^{\tau }(\boldsymbol{k}^{\prime })$
and $v_{\sigma }^{\tau }(\boldsymbol{k}^{\prime })$ are not wave functions but
envelope functions\cite{AndoReview}. The wave functions are given by $%
e^{i\tau \boldsymbol{K}\cdot \boldsymbol{x}}u_{\sigma }^{\tau }(\boldsymbol{x})$ and $%
e^{i\tau \boldsymbol{K}\cdot \boldsymbol{x}}v_{\sigma }^{\tau }(\boldsymbol{x})$.
Accordingly, the field operators of electrons and holes are 
\end{subequations}
\begin{equation}
\psi _{\text{e}\tau }(\boldsymbol{x})=e^{i\tau \boldsymbol{K}\cdot \boldsymbol{x}}\Psi _{%
\text{e}\tau }(\boldsymbol{x}),\qquad \psi _{\text{h}\tau }(\boldsymbol{x})=e^{i\tau 
\boldsymbol{K}\cdot \boldsymbol{x}}\Psi _{\text{h}\tau }(\boldsymbol{x}).
\label{RelatWE}
\end{equation}%
Thus, to analyze the Coulomb interaction, it is necessary to use $\psi _{%
\text{e}\tau }(\boldsymbol{x})$ and $\psi _{\text{h}\tau }(\boldsymbol{x})$. In
Appendix \ref{SecGraphDirac} we derive the Dirac Hamiltonian (\ref%
{DiracHamilD}) based on the effective-mass description\cite{AndoReview}.

We make a comment on the factor $\tau $ in front of $\sigma _{x}$ in (\ref%
{DiracHamilD}). It has the standard expression of the Dirac Hamiltonian for $%
\tau =+$, that is, at the K point. The K point is transformed into the K'
point under the mirror reflection. Corresponding to this we have the
mirror-reflected Dirac Hamiltonian (\ref{DiracHamilD}) for $\tau =-$, that
is, at the K' point. The total Hamiltonian (\ref{GraphHamilD}) has the
mirror symmetry.

\subsection{Landau Levels}

We apply the magnetic field to a graphene sheet taken on the $xy$ plane. It
is introduced to the Hamiltonian (\ref{GraphHamilD}) by making the minimal
substitution,%
\begin{equation}
H_{\text{D}}^{\tau }=v_{\text{F}}\left( \tau \sigma _{x}P_{x}+\sigma
_{y}P_{y}\right) \gamma _{5},  \label{DiracHamilE}
\end{equation}%
where $P_{i}\equiv -i\hbar \partial _{i}+eA_{i}$ is the covariant momentum.
We assume a homogeneous magnetic field $\boldsymbol{B}=\boldsymbol{\nabla }\times 
\boldsymbol{A}=(0,0,-B)$ with $B>0$ along the $z$ axis. The presence of the
external magnetic field modifies the mirror symmetry as follows: The
modified mirror reflection not only transforms the K point into the K' point
but also reverses the direction of the magnetic field to maintain the
symmetry. This is most clearly seen in (\ref{PauliHamil}), as we shall
discuss later. Consequently, the K and K' points become physically
distinguishable.

Electrons make cyclotron motion in magnetic field. The helicity is no longer
a good variable, since there exists a special direction for the spin, that
is, the direction of magnetic field. In this case it is convenient to use
the standard representation for the Dirac matrices, where%
\begin{equation}
\gamma _{5}=\left( 
\begin{array}{cc}
0 & 1 \\ 
1 & 0%
\end{array}%
\right) .  \label{MatriAlpha}
\end{equation}%
In order to discuss the QHE, we solve the quantum mechanical problem with
the Hamiltonian (\ref{DiracHamilE}) in each Dirac valley,%
\begin{equation}
H_{\text{D}}^{\tau }\Psi _{\tau }(\boldsymbol{x})=\mathcal{E}_{\tau }\Psi _{\tau
}(\boldsymbol{x}),  \label{G-DiracB}
\end{equation}%
and we expand the field operator in terms of the new eigenfunctions.

To solve the eigen equation (\ref{G-DiracB}) we express the Hamiltonian (\ref%
{DiracHamilE}) as%
\begin{equation}
H_{\text{D}}^{\tau }=\left( 
\begin{array}{cc}
0 & Q_{\tau } \\ 
Q_{\tau } & 0%
\end{array}%
\right) ,  \label{DiracHamilB}
\end{equation}%
with the use of (\ref{MatriAlpha}) for $\gamma _{5}$, where 
\begin{equation}
Q_{\tau }=v_{\text{F}}\left( \tau \sigma _{x}P_{x}+\sigma _{y}P_{y}\right) .
\label{ChargSUSYQ}
\end{equation}%
It is diagonalized as%
\begin{equation}
H_{\text{D}}^{\tau }=\text{diag.}\left( \sqrt{Q_{\tau }Q_{\tau }},-\sqrt{%
Q_{\tau }Q_{\tau }}\right) .  \label{DiracHamilC}
\end{equation}%
We introduce a pair of operators 
\begin{equation}
a=\frac{\ell _{B}(P_{x}+iP_{y})}{\sqrt{2}\hbar },\quad a^{\dagger }=\frac{%
\ell _{B}(P_{x}-iP_{y})}{\sqrt{2}\hbar },  \label{G-OperaA}
\end{equation}%
satisfying $[a,a^{\dag }]=1$, where $\ell _{B}=\sqrt{\hbar /eB}$ is the
magnetic length. Since the operators (\ref{ChargSUSYQ}) is rewritten as%
\begin{equation}
Q_{+}=\hbar \omega _{c}\left( 
\begin{array}{cc}
0 & a^{\dagger } \\ 
a & 0%
\end{array}%
\right) ,\quad Q_{-}=\hbar \omega _{c}\left( 
\begin{array}{cc}
0 & a \\ 
a^{\dagger } & 0%
\end{array}%
\right) ,  \label{SuperQa}
\end{equation}%
with $\omega _{c}=\sqrt{2}\hbar v_{\text{F}}/\ell _{B}$, the diagonalized
Hamiltonian reads
\begin{subequations}
\begin{align}
H_{\text{D}}^{+}=& \hbar \omega _{c}\,\text{diag.}\left( \sqrt{a^{\dagger }a}%
,\sqrt{aa^{\dagger }},-\sqrt{a^{\dagger }a},-\sqrt{aa^{\dagger }}\right) , \\
H_{\text{D}}^{-}=& \hbar \omega _{c}\,\text{diag.}\left( \sqrt{aa^{\dagger }}%
,\sqrt{a^{\dagger }a},-\sqrt{aa^{\dagger }},-\sqrt{a^{\dagger }a}\right) .
\end{align}%
Since $a^{\dagger }a$ is the number operator, the eigenvalue of the Dirac
Hamiltonian $H_{\text{D}}^{\pm }$ follows immediately,
\end{subequations}
\begin{subequations}
\label{G-LandaSpect}
\begin{align}
\mathcal{E}_{N}^{+}& =\hbar \omega _{c}\left( \sqrt{N},\sqrt{N+1},-\sqrt{N},-%
\sqrt{N+1}\right) , \\
\mathcal{E}_{N}^{-}& =\hbar \omega _{c}\left( \sqrt{N+1},\sqrt{N},-\sqrt{N+1}%
,-\sqrt{N}\right) ,
\end{align}%
with the eigenstate 
\end{subequations}
\begin{equation}
|N\rangle =\frac{1}{\sqrt{N!}}(a^{\dagger })^{N}|0\rangle ,
\end{equation}
where $N=0,1,2,3,\cdots $.

The operators $a$ and $a^{\dag }$ are the Landau-level ladder operators.
Hence, $N$ in (\ref{G-LandaSpect}) represents the Landau-level index. We
have found that the energy of an electron in the $N$th Landau level is
either $\hbar \omega _{c}\sqrt{N}$ or $\hbar \omega _{c}\sqrt{N+1}$. It is
curious that the energy of an electron in the lowest Landau level ($N=0$) is
zero though it performs cyclotron motion. This puzzle is solved in the
succeeding subsection.


\subsection{Pauli Hamiltonian}

To reveal the intrinsic structure of the energy spectrum, we investigate the
Hamiltonian\cite{Ezawa,Feynman58,Witten,Thaller92}%
\begin{equation}
H_{\text{P}}^{\pm }=Q_{\pm }Q_{\pm }=v_{\text{F}}^{2}\left[ \left( -i\hbar
\nabla +e\boldsymbol{A}\right) ^{2}\mp e\hbar \sigma _{z}B\right] ,
\label{PauliHamil}
\end{equation}%
which is the building blocks of the Dirac Hamiltonian (\ref{DiracHamilC}).
Here, the direction of the magnetic field is found to be effectively
opposite at the K and K' points. Since this has the same form as the Pauli
Hamiltonian with the mass $m^{\ast }=1/4v_{\text{F}}^{2}$ except for the
dimension, we call it the Pauli Hamiltonian for brevity. The salient feature
of the \textit{relativistic} Dirac Hamiltonian is that its spectrum is
mapped from that of the \textit{nonrelativistic} Pauli Hamiltonian. Thus,
the energy eigenvalue $\mathcal{E}_{N}^{\tau }$ of the Dirac Hamiltonian $H_{%
\text{D}}^{\tau }$ is constructed as $\mathcal{E}_{N}^{\tau }=\pm \sqrt{%
E_{N}^{\tau }}$ from the energy eigenvalue $E_{N}^{\tau }$ of the Pauli
Hamiltonian $H_{\text{P}}^{\tau }$.

In the Pauli Hamiltonian (\ref{PauliHamil}), the first term is the kinetic
term while the second term is the Zeeman term. It is fixed uniquely as an
intrinsic property of the Dirac theory: We call it the intrinsic Zeeman
effect.

The Landau level is created by electrons making cyclotron motion. In the
conventional QHE, since the Zeeman energy can be considered much smaller
than the Landau-level separation, we may treat it as a perturbation.
However, this is not the case in graphene. According to the Pauli
Hamiltonian (\ref{PauliHamil}), the intrinsic Zeeman energy is precisely one
half of the cyclotron energy for Dirac electrons, and two Landau levels mix
to create one energy level, as illustrated in Fig.\ref{FigGraphLevel}(a).

We consider the K point ($\tau =+$). It is obvious that the up-spin and
down-spin states are eigenstates of the Pauli Hamiltonian (\ref{PauliHamil})
and hence eigenstates of the Dirac Hamiltonian (\ref{DiracHamilC}), and that
the up-spin state has a lower energy than the down-spin state when they
belong to the same Landau level. On the other hand, the direction of the
spin is opposite at the K and K' points. Hence, we can make the following
identification of quantum numbers for electrons in the $N$th level ($N\geq 1$%
) of the energy spectrum [Fig.\ref{FigGraphLevel}(a)],%
\begin{equation}
\mathcal{E}_{N}^{+\uparrow }=\mathcal{E}_{N-1}^{+\downarrow }=\mathcal{E}%
_{N}^{-\downarrow }=\mathcal{E}_{N-1}^{-\uparrow }=\hbar \omega _{c}\sqrt{N}.
\label{G-EnergSpect}
\end{equation}%
A similar identification can be made for holes in the $N$th energy level ($%
N\geq 1$) . The zeroth energy level ($N=0$) consists of the up-spin electron
and the down-spin hole at the K point, and the down-spin electron and the
up-spin hole at the K' point [Fig.\ref{FigGraphLevel}(a)], coming from the
lowest Landau levels for electrons and holes.

Counting the states at the K and K' points all together, one energy level
has a 4-fold degeneracy. Each filled energy level contributes one
conductance quantum $e^{2}/\hbar $ to the Hall conductivity. Consequently
the resulting series is $\nu =\pm 2,\pm 6,\pm 10,\cdots $, which accounts
for the bold-face series (\ref{G-SerieMono}) in the monolayer graphene.

\subsection{Field Operators}

We focus on electrons in the $N$th energy level. (Essentially the same
analysis is applicable to holes.) We start with the quantum-mechanical state
in a single Landau level. We decompose the electron coordinate $\boldsymbol{x}%
=(x,y)$ into the guiding center $\boldsymbol{X}=(X,Y)$ and the relative
coordinate $\boldsymbol{R}=(R_{x},R_{y})$, $\boldsymbol{x}=\boldsymbol{X}+\boldsymbol{R}$,
where $R_{x}=-P_{y}/eB$ and $R_{y}=P_{x}/eB$ with $\boldsymbol{P}=(P_{x},P_{y})$
the covariant momentum. They satisfy the commutation relations $[X,Y]=-i\ell
_{B}^{2}$, $[P_{x},P_{y}]=i{\hbar ^{2}/\ell _{B}^{2}}$, $%
[X,P_{x}]=[X,P_{y}]=[Y,P_{x}]=[Y,P_{y}]=0$. We define a set of operators $b$%
, $b^{\dag }$ by%
\begin{equation}
b=\frac{1}{\sqrt{2}\ell _{B}}(X-iY),\quad b^{\dag }=\frac{1}{\sqrt{2}\ell
_{B}}(X+iY),
\end{equation}%
obeying $[b,b^{\dag }]=1$, in addition to a set of operators $a$, $a^{\dag }$
by (\ref{G-OperaA}). The quantum-mechanical states are the Fock states, 
\begin{equation}
|N,n\rangle =\frac{1}{\sqrt{N!n!}}(a^{\dag })^{N}(b^{\dag })^{n}|0\rangle
\label{G-FockNn}
\end{equation}%
with $|0\rangle $ the Fock vacuum, $a|0\rangle =b|0\rangle =0$.

We have constructed the energy spectrum of the Dirac Hamiltonian (\ref%
{DiracHamilC}) in each Dirac valley, as illustrated in Fig.\ref%
{FigGraphLevel}(a). The $N$th energy level ($N\neq 0$) contains electrons
coming from two Dirac valleys ($\tau =\pm $), whose field operators are
expanded in terms of the eigenfunctions. Corresponding to $\mathcal{E}%
_{N}^{+\uparrow }$, $\mathcal{E}_{N-1}^{+\downarrow }$, $\mathcal{E}%
_{N}^{-\downarrow }$ and $\mathcal{E}_{N-1}^{-\uparrow }$ in (\ref%
{G-EnergSpect}), we have
\begin{subequations}
\label{G-ProjeElect}
\begin{align}
\psi _{N+}^{\uparrow }(\boldsymbol{x})& =e^{i\boldsymbol{Kx}}\sum_{n}\langle \boldsymbol{%
x}|N,n\rangle c_{+}^{\uparrow }(N,n), \\
\psi _{N-}^{\downarrow }(\boldsymbol{x})& =e^{-i\boldsymbol{Kx}}\sum_{n}\langle 
\boldsymbol{x}|N,n\rangle c_{-}^{\downarrow }(N,n), \\
\psi _{N+}^{\downarrow }(\boldsymbol{x})& =e^{i\boldsymbol{Kx}}\sum_{n}\langle 
\boldsymbol{x}|N-1,n\rangle c_{+}^{\downarrow }(N-1,n), \\
\psi _{N-}^{\uparrow }(\boldsymbol{x})& =e^{-i\boldsymbol{Kx}}\sum_{n}\langle 
\boldsymbol{x}|N-1,n\rangle c_{-}^{\uparrow }(N-1,n),
\end{align}%
where $c_{\tau }^{\sigma }(N,n)$ is the annihilation operator acting on the
Fock state $|N,n\rangle $ at the $\tau $ point. In what follows we suppress
the Landau-level index $N$ in $c_{\tau }^{\sigma }(N,n)$ with the
understanding that $c_{+}^{\uparrow }(n)=c_{+}^{\uparrow }(N,n)$, $%
c_{-}^{\downarrow }(n)=c_{-}^{\downarrow }(N,n)$, $c_{+}^{\downarrow
}(n)=c_{+}^{\downarrow }(N-1,n)$, $c_{-}^{\uparrow }(n)=c_{-}^{\uparrow
}(N-1,n)$. See (\ref{RelatWE}) with respect to the factor $e^{\pm i\boldsymbol{Kx%
}}$. The electron operators in the zeroth energy level are given by $\psi
_{N+}^{\uparrow }(\boldsymbol{x})$ and $\psi _{N-}^{\downarrow }(\boldsymbol{x})$
with $N=0$.

\section{Coulomb Interactions}

\label{SecGraphCoulo}

We make the basic assumption that the cyclotron energy is much larger than
the Coulomb energy as in the conventional QHE\cite%
{BookPrange,BookDasSarma,BookEzawa}. We neglect the Landau-level mixing by
the Coulomb interaction, and analyze it within one energy level [Fig.\ref%
{FigGraphLevel}(a)]. There exists a consistent formalism, known as the
Landau-level projection. The projected theory presents not only a good
approximation but also an essential way to reveal a new physics inherent to
the QH system\cite{BookPrange,BookDasSarma,BookEzawa}. Then we may treat the
electron system, the hole system and the zero-energy system independently of
each other. The hole system has the same structure as the electron system
due to the electron-hole symmetry. On the other hand, the zero-energy system
contains both electrons and holes. We study the electron system in this
section, and the zero-energy system of electrons and holes in Section \ref%
{SecExciton}.

Assuming all lower levels are filled up, we study the Coulomb interaction
between electrons confined within the $N$th energy level ($N\geq 1$). The
Coulomb Hamiltonian reads
\end{subequations}
\begin{equation}
H=\frac{1}{2}\int \int \!d^{2}xd^{2}y\,V(\boldsymbol{x}-\boldsymbol{y})\rho (\boldsymbol{%
x})\rho (\boldsymbol{y}),  \label{G-CouloHamil}
\end{equation}%
where $\rho \left( \boldsymbol{x}\right) $ is the density operator 
\begin{equation}
\rho (\boldsymbol{x})=\sum_{\tau \tau ^{\prime }\sigma }\psi _{\tau }^{\sigma
\dag }\left( \boldsymbol{x}\right) \psi _{\tau ^{\prime }}^{\sigma }\left( 
\boldsymbol{x}\right) .  \label{G-densi}
\end{equation}%
The Landau-level projection of the Coulomb Hamiltonian (\ref{G-CouloHamil})
is a simple generalization of the lowest-Landau-level projection\cite%
{Girvin84B} familiar in the conventional QHE. We require the density
operator to be comprised solely of the electron fields belonging to the $N$%
th energy level. Thus, from (\ref{G-densi}) we construct the projected
density operator $\rho _{N}(\boldsymbol{x})$ by 
\begin{equation}
\rho _{N}(\boldsymbol{x})=\sum_{\tau \tau ^{\prime }\sigma }\psi _{N\tau
}^{\sigma \dag }(\boldsymbol{x})\psi _{N\tau ^{\prime }}^{\sigma }(\boldsymbol{x}),
\end{equation}%
where $\psi _{N\tau }^{\sigma }(\boldsymbol{x})$ is the field operator (\ref%
{G-ProjeElect}). As we show in Appendix \ref{AppenLLProje}, the projected
density operator is rewritten in the momentum space as%
\begin{equation}
\rho _{N}(\boldsymbol{q})=\sum_{\tau \tau ^{\prime }\sigma }F_{\tau \tau
^{\prime }}^{\sigma }\left( \boldsymbol{q}\right) \hat{D}_{\tau \tau ^{\prime
}}^{\sigma \sigma }\left( \boldsymbol{q}\right) ,  \label{DensiInN}
\end{equation}%
where $\hat{D}_{\tau \tau ^{\prime }}^{\sigma \sigma ^{\prime }}(\boldsymbol{q})$
is the bare density operator\cite{Ezawa05D},%
\begin{equation}
\hat{D}_{\tau \tau ^{\prime }}^{\sigma \sigma ^{\prime }}(\boldsymbol{q})=\frac{1%
}{2\pi }\sum_{mn}\langle m|e^{-i[\boldsymbol{q}+\tau \boldsymbol{K}-\tau ^{\prime }%
\boldsymbol{K}]\boldsymbol{X}}|n\rangle c_{\tau }^{\sigma \dag }(m)c_{\tau ^{\prime
}}^{\sigma ^{\prime }}(n),  \label{G-BareDensi}
\end{equation}%
and $F_{\tau \tau ^{\prime }}^{\sigma }\left( \boldsymbol{q}\right) $ is the
form factor,
\begin{subequations}
\label{G-FormFactor}
\begin{align}
F_{++}^{\uparrow }\left( \boldsymbol{q}\right) & =F_{--}^{\downarrow }\left( 
\boldsymbol{q}\right) =F_{N}\left( \boldsymbol{q}\right) , \\
F_{++}^{\downarrow }\left( \boldsymbol{q}\right) & =F_{--}^{\uparrow }\left( 
\boldsymbol{q}\right) =F_{N-1}\left( \boldsymbol{q}\right) , \\
F_{+-}^{\uparrow }\left( \boldsymbol{q}\right) & =F_{-+}^{\downarrow }\left( 
\boldsymbol{q}\right) =G_{N}\left( \boldsymbol{q}\right) , \\
F_{-+}^{\uparrow }\left( \boldsymbol{q}\right) & =F_{+-}^{\downarrow }\left( 
\boldsymbol{q}\right) =G_{N}^{\ast }\left( -\boldsymbol{q}\right) ,
\end{align}%
with
\end{subequations}
\begin{subequations}
\begin{align}
F_{N}\left( \boldsymbol{q}\right) =& \langle N|e^{-i\boldsymbol{qR}}|N\rangle , \\
G_{N}\left( \boldsymbol{q}\right) =& \langle N|e^{-i(\boldsymbol{q}-\boldsymbol{K)R}%
}|N-1\rangle .
\end{align}%
Here we have used the relation, $\boldsymbol{q}=\boldsymbol{q}+3\boldsymbol{K}$, due to
the lattice structure. It should be remarked that the bare density operator $%
\hat{D}_{\tau \tau ^{\prime }}^{\sigma \sigma ^{\prime }}(\boldsymbol{q})$
involves only the guiding center coordinate $\boldsymbol{X}$, while the form
factor $F_{\tau \tau ^{\prime }}^{\sigma }\left( \boldsymbol{q}\right) $
involves only the relative coordinate $\boldsymbol{R}$.

The form factors are explicitly given by using\cite{Ando74JPSJ}
\end{subequations}
\begin{equation}
\langle N\text{+}M|e^{i\boldsymbol{qR}}|N\rangle =\frac{\sqrt{N!}}{\sqrt{(N+M)!}}%
\left( \frac{\ell _{B}q}{\sqrt{2}}\right) ^{M}L_{N}^{M}\left( \frac{\ell
_{B}^{2}\boldsymbol{q}^{2}}{2}\right) e^{-\frac{1}{4}\ell _{B}^{2}\boldsymbol{q}^{2}}
\notag
\end{equation}%
for $M\geq 0$ in terms of associated Laguerre polynomials.

A comment is in order on the form factors. It is easy to see that\cite%
{Goerbig} 
\begin{equation}
G_{N}\left( \boldsymbol{q}\right) \simeq e^{-\frac{1}{2}\ell _{B}^{2}|\boldsymbol{K|}%
^{2}}F_{N}\left( \boldsymbol{q}\right) .  \label{G-FactoK}
\end{equation}%
Thus, $G_{N}\left( \boldsymbol{q}\right) $ is exponentially smaller than $%
F_{N}\left( \boldsymbol{q}\right) $. For instance, $F_{+-}^{\uparrow }\left( 
\boldsymbol{q}\right) \equiv G_{N}\left( \boldsymbol{q}\right) $ represents the
transfer of the up-spin electron ($\sigma =\uparrow $) from the K' point ($%
\tau =-$) to the K point ($\tau =+$). Hence, such a mixing between the K and
K' points is actually negligible.

\begin{figure}[t]
\begin{center}
\includegraphics[width=0.55\textwidth]{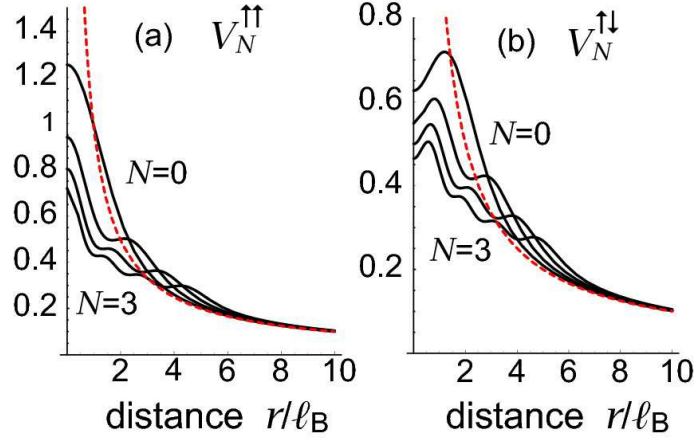}
\end{center}
\caption{\baselineskip=12pt (Color online) The spin dependence of the effective Coulomb
potential. The vertical axis is the potential in unit of $e^{2}/4\protect\pi 
\protect\varepsilon \ell _{B}$. The horizontal axis is the distance $r$ in
unit of $\ell _{B}$. The dotted red curve represents the ordinary Coulomb
potential without the form factor. (a) The effective Coulomb potential $%
V_{N}^{\uparrow \uparrow }(r)$ in the $N$th energy level for $N=0,1,2,3$
from top to bottom. Note that $V_{N}^{\downarrow \downarrow
}(r)=V_{N-1}^{\uparrow \uparrow }(r)$. It describes interactions between
electrons with the same spin. (b) The effective Coulomb potential $%
V_{N}^{\uparrow \downarrow }(r)$ for $N=0,1,2,3$ from top to bottom. It
describes interactions between electrons with the different spins. }
\label{FigProjeCoulo}
\end{figure}

The projected Coulomb Hamiltonian is constructed by substituting the
projected density (\ref{DensiInN}) into the Hamiltonian (\ref{G-CouloHamil}%
), 
\begin{align}
H_{N}=& \pi \int {\!d^{2}q\,}V(\boldsymbol{q})\rho _{N}(-\boldsymbol{q})\rho _{N}(%
\boldsymbol{q})  \notag \\
=& \pi \sum_{\tau \tau ^{\prime }\sigma ^{\prime }}\sum_{\lambda \lambda
^{\prime }\sigma ^{\prime }}\int {\!d^{2}q\,}V_{N;\tau \tau ^{\prime
}\lambda \lambda ^{\prime }}^{\sigma \sigma ^{\prime }}(\boldsymbol{q})\hat{D}%
_{\tau \tau ^{\prime }}^{\sigma }(-\boldsymbol{q})\hat{D}_{\lambda \lambda
^{\prime }}^{\sigma ^{\prime }}(\boldsymbol{q}),  \label{ProjeHamil}
\end{align}%
where $V_{N;\tau \tau ^{\prime }\lambda \lambda ^{\prime }}^{\sigma \sigma
^{\prime }}(\boldsymbol{q})$ is the effective Coulomb potential in the $N$th
energy level,%
\begin{equation}
V_{N;\tau \tau ^{\prime }\lambda \lambda ^{\prime }}^{\sigma \sigma ^{\prime
}}(\boldsymbol{q})=V(\boldsymbol{q})F_{\tau \tau ^{\prime }}^{\sigma }(-\boldsymbol{q}%
)F_{\lambda \lambda ^{\prime }}^{\sigma ^{\prime }}(\boldsymbol{q})
\label{ProjeCouloPoten}
\end{equation}%
with%
\begin{equation}
V(\boldsymbol{q})=\frac{e^{2}}{4\pi \varepsilon |\boldsymbol{q}|}.
\end{equation}%
It is remarkable that the effective Coulomb potential depends on the spin
and the valley through the form factors $F_{\tau \tau ^{\prime }}^{\sigma }(%
\boldsymbol{q})$ characterizing Landau levels.

The typical effective Coulomb potentials in the $N$th energy level at the K
point are
\begin{subequations}
\label{G-EffecPoten}
\begin{align}
V_{N}^{\uparrow \uparrow }\left( \boldsymbol{q}\right) =& V(\boldsymbol{q})F_{N}(-%
\boldsymbol{q})F_{N}(\boldsymbol{q}), \\
V_{N}^{\downarrow \downarrow }\left( \boldsymbol{q}\right) =& V(\boldsymbol{q}%
)F_{N-1}(-\boldsymbol{q})F_{N-1}(\boldsymbol{q}), \\
V_{N}^{\uparrow \downarrow }\left( \boldsymbol{q}\right) =& V(\boldsymbol{q})F_{N}(-%
\boldsymbol{q})F_{N-1}(\boldsymbol{q}).
\end{align}%
The potential $V_{N}^{\sigma \sigma }\left( \boldsymbol{q}\right) $ stands for
the interaction between electrons with the same spin $\sigma $, $%
V_{N}^{\uparrow \downarrow }\left( \boldsymbol{q}\right) $ for the interaction
between electrons with the different spin. The spin direction is reversed at
the K' point. We have illustrated the effective Coulomb potentials in Figs.%
\ref{FigProjeCoulo} and \ref{FigProjN}.

\begin{figure}[t]
\begin{center}
\includegraphics[width=0.8\textwidth]{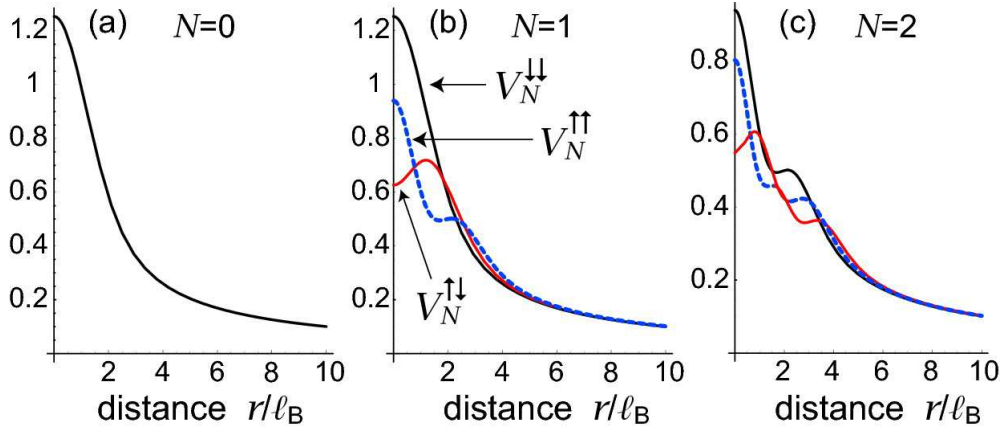}
\end{center}
\caption{\baselineskip=12pt (Color online) The spin dependence of the effective Coulomb
potential $V_{N}^{\downarrow \downarrow }(r)$, $V_{N}^{\uparrow \uparrow }(r)
$ and $V_{N}^{\uparrow \downarrow }(r)$ in the $N$th energy level for (a) $%
N=0$, (b) $N=1$, and (c) $N=2$. All effective potentials agree for the
lowest energy level ($N=0$). The vertical axis is the potential in unit of $%
e^{2}/4\protect\pi \protect\varepsilon \ell _{B}$. }
\label{FigProjN}
\end{figure}

We note that the effective potential $V_{N}^{\uparrow \uparrow }\left( 
\boldsymbol{q}\right) $ for electrons with the same spin in the $N$th energy
level has precisely the same form as in the standard QHE for electrons in
the $N$th Landau level: Compare Fig.\ref{FigProjeCoulo}(a) with Fig.7 in Ref.%
\cite{Shibata}. On the other hand the effective potential $V_{N}^{\uparrow
\downarrow }\left( \boldsymbol{q}\right) $ for electrons with the different
spins are entirely new.

\section{Excitonic Condensation}

\label{SecExciton}

It is necessary to pay a special attention to the zero-energy level ($N=0$),
since it contains both electrons and holes. Let us recapture the properties
of the zero-energy level [Fig.\ref{FigElectHoleG}]. In the absence of the
magnetic field the band structure is given by the Dirac valleys associated
with the dispersion relation (\ref{DispeDirac}), as illustrated in Fig.\ref%
{FigElectHoleG}(a). When the magnetic field is applied, the band structure
is changed to generate Landau levels [Fig.\ref{FigElectHoleG}(b)]. However,
the Dirac electron is subject to the intrinsic Zeeman effect, which induces
a Landau-level mixing, and the zero-energy state emerges for up-spin
electrons as well as down-spin holes [Fig.\ref{FigElectHoleG}(c)]. There
exists an attractive Coulomb force between an electron and a hole. Hence we
expect them to make an excitonic condensation, producing a gap to the
electron and hole states [Fig.\ref{FigElectHoleG}(d)].

\begin{figure}[t]
\begin{center}
\includegraphics[width=0.7\textwidth]{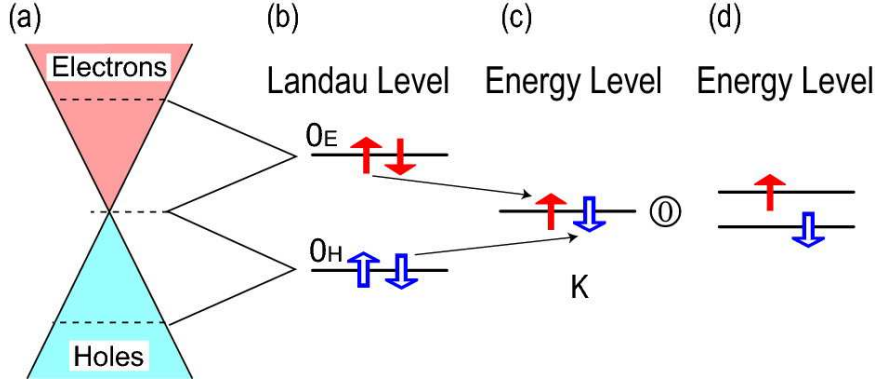}
\end{center}
\caption{\baselineskip=12pt (Color online) (a) Electron states (blue region) and hole states \
(red region) in the Dirac valley at the K point. (b) When the magnetic field
is applied, the band structure is changed to generate Landau levels. (c) The
intrinsic Zeeman effect induces a Landau-level mixing, and the zero-energy
state with the up-spin electron and the down-spin hole emerges. (d) They
make an excitonic condensation to form a BCS-type state, producing a gap to
the electron and hole states. }
\label{FigElectHoleG}
\end{figure}

The excitonic condensation in graphene has been studied in various contexts%
\cite{Khveschenko01La,Khveschenko01Lb,Gorbar02B,Gusynin06B}. Here, we
present a clear-cut approach to this problem on the analogy of the BCS
theory. Our physical picture is summarized in Fig.\ref{FigElectHoleG}. The
remarkable point is that the kinetic term is quenched in each Landau level,
and hence it is also absent in the zero-energy level. This simplifies
considerably our analysis.

The field operators present in the zero-energy level are $\psi _{\text{e}%
+}^{\uparrow }(\boldsymbol{x})$, $\psi _{\text{e}-}^{\downarrow }(\boldsymbol{x})$
for electrons and $\psi _{\text{h}+}^{\downarrow }(\boldsymbol{x})$, $\psi _{%
\text{h}-}^{\uparrow }(\boldsymbol{x})$ for holes coming from the $N=0$ Landau
levels [Fig.\ref{FigGraphLevel}(a)]. It is sufficient to investigate
electron-hole pairs at the K and K' points separately. An exciton composed
of an electron and a hole belonging to different valleys is fragile, because
their effective Coulomb potential involves the factor $e^{-\frac{1}{2}\ell
_{B}^{2}|\boldsymbol{K|}^{2}}$ compared with the one within the K point: See (%
\ref{G-FormFactor}) and (\ref{G-FactoK}). For definiteness, we consider the
K point [Fig.\ref{FigElectHoleG}].

The effective Coulomb potential (\ref{ProjeCouloPoten}) is simple in the
lowest Landau level,
\end{subequations}
\begin{equation}
V^{\text{eff}}(\boldsymbol{q})=V(\boldsymbol{q})e^{-\ell _{B}^{2}\boldsymbol{q}^{2}/2}.
\end{equation}%
We consider the effective Hamiltonian together with this effective
potential, 
\begin{equation}
H^{\text{eff}}=-\int \!d^{2}xd^{2}y\,V^{\text{eff}}(\boldsymbol{x}-\boldsymbol{y}%
)\psi _{\text{e}}^{\uparrow \dag }(\boldsymbol{x})\psi _{\text{e}}^{\uparrow }(%
\boldsymbol{x})\psi _{\text{h}}^{\downarrow \dag }(\boldsymbol{y})\psi _{\text{h}%
}^{\downarrow }(\boldsymbol{y}),  \label{G-HamilEH}
\end{equation}%
where $\psi _{\text{e}}^{\uparrow }(\boldsymbol{x})$ and $\psi _{\text{h}%
}^{\downarrow }(\boldsymbol{x})$ are the up-spin electron field and the
down-spin hole field, respectively. The Hamiltonian is rewritten as 
\begin{equation}
H^{\text{eff}}=-\pi {\!\!}\int {\!}V^{\text{eff}}(\boldsymbol{q})\psi _{\text{e}%
}^{\uparrow \dag }(\boldsymbol{k}+\boldsymbol{q})\psi _{\text{h}}^{\downarrow \dag }(%
\boldsymbol{k}^{\prime })\psi _{\text{h}}^{\downarrow }(\boldsymbol{k}^{\prime }-%
\boldsymbol{q})\psi _{\text{e}}^{\uparrow }(\boldsymbol{k}).
\end{equation}%
Here and here after, under the symbol $\int $, the integration over momentum
variables is understood.

We derive the gap equation, following the analysis familiar in the BCS
theory. We take the terms satisfying $\boldsymbol{q}=\boldsymbol{k}^{\prime }-%
\boldsymbol{k}$ as the dominant ones, and approximate the Hamiltonian as 
\begin{equation}
H^{\text{eff}}\simeq -\pi {\!\!}\int {\!}V^{\text{eff}}(\boldsymbol{k}^{\prime }-%
\boldsymbol{k)}\psi _{\text{e}}^{\uparrow \dag }(\boldsymbol{k}^{\prime })\psi _{%
\text{h}}^{\downarrow \dag }(\boldsymbol{k}^{\prime })\psi _{\text{h}%
}^{\downarrow }(\boldsymbol{k})\psi _{\text{e}}^{\uparrow }(\boldsymbol{k}).
\end{equation}%
We define the singlet excitonic gap function by%
\begin{equation}
\Delta \left( \boldsymbol{k}\right) =\int {\!}d^{2}k^{\prime }\,V^{\text{eff}}(%
\boldsymbol{k}^{\prime }-\boldsymbol{k)}\langle \psi _{\text{e}}^{\uparrow \dag }(%
\boldsymbol{k}^{\prime })\psi _{\text{h}}^{\downarrow \dag }(\boldsymbol{k}^{\prime
})\rangle ,  \label{GapEquat}
\end{equation}%
which can be taken to be positive without loss of generality.

The mean-field Hamiltonian reads%
\begin{equation}
H^{\text{eff}}\simeq -\pi {\!\!}\int {\![}\Delta \left( \boldsymbol{k}\right)
\psi _{\text{h}}^{\downarrow }\left( \boldsymbol{k}\right) \psi _{\text{e}%
}^{\uparrow }\left( \boldsymbol{k}\right) +\psi _{\text{e}}^{\uparrow \dag
}\left( \boldsymbol{k}\right) \psi _{\text{h}}^{\downarrow \dag }\left( \boldsymbol{k%
}\right) \Delta ^{\ast }\left( \boldsymbol{k}\right) ].  \label{G-HamilEHx}
\end{equation}%
By setting
\begin{subequations}
\label{G-StepB}
\begin{align}
\Psi _{1}\left( \boldsymbol{k}\right) =& \frac{1}{\sqrt{2}}[\psi _{\text{e}%
}^{\uparrow }\left( \boldsymbol{k}\right) +\psi _{\text{h}}^{\downarrow \dag
}\left( \boldsymbol{k}\right) ], \\
\Psi _{2}\left( \boldsymbol{k}\right) =& \frac{1}{\sqrt{2}}[\psi _{\text{e}%
}^{\uparrow \dagger }\left( \boldsymbol{k}\right) -\psi _{\text{h}}^{\downarrow
}\left( \boldsymbol{k}\right) ]2,
\end{align}%
it is easy to diagonalize (\ref{G-HamilEHx}) as 
\end{subequations}
\begin{equation}
H^{\text{eff}}\simeq \pi \int d^{2}k\,\Delta \left( \boldsymbol{k}\right) [\Psi
_{1}^{\dagger }\left( \boldsymbol{k}\right) \Psi _{1}\left( \boldsymbol{k}\right)
+\Psi _{2}^{\dagger }\left( \boldsymbol{k}\right) \Psi _{2}\left( \boldsymbol{k}%
\right) ].  \label{G-DispeAB}
\end{equation}%
Since $\Psi _{1}$ and $\Psi _{2}$ are free fields, the ground state is given
by solving%
\begin{equation}
\Psi _{1}\left( \boldsymbol{k}\right) |\Phi _{\text{exc}}\rangle =\Psi
_{2}\left( \boldsymbol{k}\right) |\Phi _{\text{exc}}\rangle =0,
\end{equation}%
or%
\begin{equation}
|\Phi _{\text{exc}}\rangle =\prod\limits_{\boldsymbol{k}}\frac{1}{\sqrt{2}}%
[1+\psi _{\text{h}}^{\downarrow \dag }\left( \boldsymbol{k}\right) \psi _{\text{e%
}}^{\uparrow \dag }\left( \boldsymbol{k}\right) ]|0\rangle .
\end{equation}%
This is a BCS-type state representing the condensation of electron-hole
pairs.

Due to the Fermi statistics the thermodynamical average $\langle \Psi
_{i}^{\dagger }\left( \boldsymbol{k}\right) \Psi _{j}\left( \boldsymbol{k}\right)
\rangle $ is given by%
\begin{align}
\langle \Psi _{1}^{\dagger }\left( \boldsymbol{k}\right) \Psi _{1}\left( \boldsymbol{%
k}\right) \rangle =& \langle \Psi _{2}^{\dagger }\left( \boldsymbol{k}\right)
\Psi _{2}\left( \boldsymbol{k}\right) \rangle =\frac{1}{1+e^{\Delta \left( 
\boldsymbol{k}\right) /k_{\text{B}}T}},  \notag \\
\langle \Psi _{2}^{\dagger }\left( \boldsymbol{k}\right) \Psi _{1}\left( \boldsymbol{%
k}\right) \rangle =& \langle \Psi _{1}^{\dagger }\left( \boldsymbol{k}\right)
\Psi _{2}\left( \boldsymbol{k}\right) \rangle =0,
\end{align}%
where $k_{\text{B}}$ is the Boltzmann factor. Combining these with (\ref%
{G-StepB}) we obtain%
\begin{equation}
\langle \psi _{\text{e}}^{\uparrow \dag }\left( \boldsymbol{k}\right) \psi _{%
\text{h}}^{\downarrow \dag }\left( \boldsymbol{k}\right) \rangle =\frac{1}{2}%
\tanh \frac{\Delta \left( \boldsymbol{k}\right) }{2k_{\text{B}}T}.
\end{equation}%
Substituting this into (\ref{GapEquat}) we derive the gap equation,

\begin{equation}
\Delta \left( \boldsymbol{k}\right) =\frac{1}{2}\int d^{2}\boldsymbol{k}^{\prime }V^{%
\text{eff}}(\boldsymbol{k}^{\prime }-\boldsymbol{k)}\tanh \frac{\Delta (\boldsymbol{k}%
^{\prime })}{2k_{\text{B}}T}.
\end{equation}%
In the limit $T\rightarrow 0$, the zero-momentum gap $\Delta \left( \boldsymbol{k%
}\right) $ is given by the dispersionless relation%
\begin{equation}
\left. \Delta \left( \boldsymbol{k}\right) \right\vert _{T=0}=\pi \sqrt{\frac{%
\pi }{2}}\frac{e^{2}}{4\pi \varepsilon \ell _{B}}\equiv \Delta _{0}.
\label{GapAtZero}
\end{equation}%
For finite temperature, assuming the gap is dispersionless, we obtain the
relation%
\begin{equation}
\frac{\Delta _{T}}{\Delta _{0}}=\tanh \frac{\Delta _{T}}{2k_{\text{B}}T}.
\end{equation}%
The critical temperature $T_{\text{C}}$ at which $\Delta _{T}=0$ is solved
as $T_{\text{C}}=\Delta _{0}/2k_{\text{B}}$.

\begin{figure}[t]
\begin{center}
\includegraphics[width=0.8\textwidth]{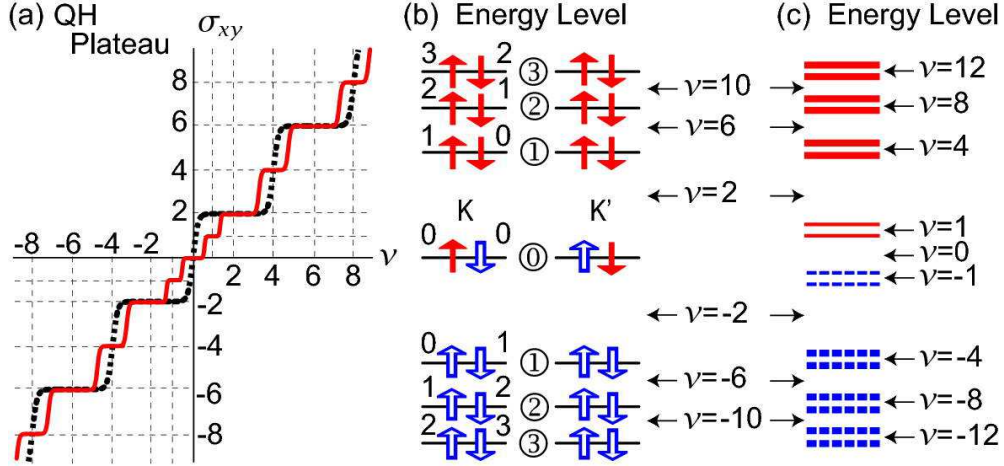}
\end{center}
\caption{\baselineskip=12pt{}(Color online) (a) The QH conductivity in graphene. The dotted
black curve shows the sequence $\protect\nu =\pm 2,\pm 6,\pm 10,\cdots $,
while the solid red curve the sequence $\protect\nu =0,\pm 1,\pm 2,\pm 4,\pm
6,\pm 8,\cdots $. (b) The energy level within noninteracting theory,
indicated by the number in circle. The spin is indicated by a solid red
(open blue) arrow for electron (hole) at the K and K' points. The number
attached to a solid red (open blue) arrow shows from which Landau level the
electron (hole) comes from. (c) When Coulomb interactions are included, the
zero-energy level splits into four nondegenerate subbands, while each
nonzero-energy level splits into two 2-fold degenerate subbands with the
exact U(1) symmetry. }
\label{FigSplit}
\end{figure}

According to the diagonalized Hamiltonian (\ref{G-DispeAB}), the excitonic
condensation provides them with the mass $\Delta _{0}$, resolving the
electron-hole degeneracy in the zero-energy state. Combining the results at
the K and K' points, the 4-fold degenerate levels split into two 2-fold
degenerate levels [Fig.\ref{FigElectHoleG}(d)] with the gap energy (\ref%
{GapAtZero}). This leads to a new plateau at $\nu =0$, as illustrated in Fig.%
\ref{FigSplit}

\section{Valley Polarization}

\label{SecKKasymm}

It is now possible to treat electrons and holes separately since the gap has
opened between the electron and hole bands. Furthermore, it is enough to
study only electrons due to the electron-hole symmetry. We show that the
Coulomb effect modifies the energy spectrum of the noninteracting theory so
that plateaux emerges at $\nu =\pm 1,\pm 2n$ with $n=1,2,3,\cdots $ [Fig.\ref%
{FigSplit}].

\begin{figure}[t]
\begin{center}
\includegraphics[width=0.8\textwidth]{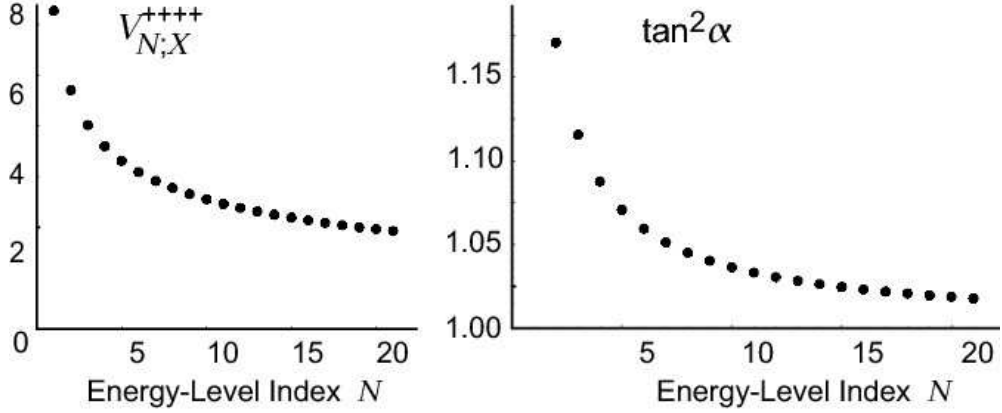}
\end{center}
\caption{\baselineskip=12pt (a) The exchange energy $V_{N;\text{X}}^{++++}$ of the $N$th energy
level as a function of $N$. The vertical axis is the energy in unit of $%
e^{2}/4\protect\pi \protect\varepsilon \ell _{B}$. Note that $V_{N;\text{X}%
}^{----}=V_{N-1;\text{X}}^{++++}$. (b) The function $\tan ^{2}\protect\alpha %
_{N}$ $\ $given by (\protect\ref{Tan}). It approaches $1$ assymptotically, $%
\lim_{N\rightarrow \infty }\tan ^{2}\protect\alpha _{N}=1$. }
\label{FigProjV}
\end{figure}

We express the Coulomb Hamiltonian (\ref{ProjeHamil}) as%
\begin{equation}
H_{N}=\sum V_{N;mnij}^{\sigma \sigma ^{\prime };\tau \tau ^{\prime }\lambda
\lambda ^{\prime }}c_{\tau }^{\sigma \dag }(m)c_{\tau ^{\prime }}^{\sigma
}(n)c_{\lambda }^{\sigma ^{\prime }\dag }(i)c_{\lambda ^{\prime }}^{\sigma
^{\prime }}(j),  \label{ProjeHamilA}
\end{equation}%
where the summations over repeated indices, $m$, $n$, $i$, $j$, $\sigma $, $%
\sigma ^{\prime }$, $\tau $, $\tau ^{\prime }$, $\lambda $, $\lambda
^{\prime }$ are understood, and%
\begin{align}
V_{N;mnij}^{\sigma \sigma ^{\prime };\tau \tau ^{\prime }\lambda \lambda
^{\prime }}=& \frac{1}{4\pi }\int {\!d^{2}q\,}V(\boldsymbol{q})F_{\tau \tau
^{\prime }}^{\sigma }(-\boldsymbol{q})F_{\lambda \lambda ^{\prime }}^{\sigma
^{\prime }}(\boldsymbol{q})  \notag \\
& \times \langle m|e^{-i[-\boldsymbol{q}+\tau \boldsymbol{K}-\tau ^{\prime }\boldsymbol{K%
}]\boldsymbol{X}}|n\rangle \langle i|e^{-i[\boldsymbol{q}+\lambda \boldsymbol{\boldsymbol{K}}%
-\lambda ^{\prime }\boldsymbol{K}]\boldsymbol{X}}|j\rangle .
\end{align}%
We introduce%
\begin{equation}
V_{N;\text{D}}^{\tau \tau ^{\prime }\lambda \lambda ^{\prime
}}=\sum_{ij}V_{N;iijj}^{\uparrow \uparrow ;\tau \tau ^{\prime }\lambda
\lambda ^{\prime }},\quad V_{N;\text{X}}^{\tau \tau ^{\prime }\lambda
\lambda ^{\prime }}=\sum_{ij}V_{N;ijji}^{\uparrow \uparrow ;\tau \tau
^{\prime }\lambda \lambda ^{\prime }}.  \label{G-VDX}
\end{equation}%
They represent the direct and exchange Coulomb energies, respectively. It is
easy to evaluate them numerically: See Fig.\ref{FigProjV}(a) for $V_{N;\text{%
X}}^{++++}$. We analyze the lowest energy level and higher energy levels
separately.

\subsection{Lowest Energy Level}

We first study the lowest energy level, which contain two degenerate states
[Fig.\ref{FigSplit}(b)]. The degeneracy is resolved obviously by a strong
extrinsic Zeeman effect if it exists. We consider the case where the
extrinsic Zeeman energy is absent at temperature $T=0$, or is present but
quite small compared with the thermal energy at $T\neq 0$. We ask whether
the degeneracy is resolved even in such cases.

As a generic trial function we take%
\begin{equation}
|\Phi _{0}\rangle =\prod\limits_{n}\left( uc_{+}^{\uparrow \dagger }\left(
n\right) +vc_{-}^{\downarrow \dagger }\left( n\right) \right) \left\vert
0\right\rangle ,
\end{equation}%
with $|u|^{2}+|v|^{2}=1$. The state covers the entire SU(2) space of the
lowest energy level for electrons, when the two parameters $u$ and $v$ are
varied.

We calculate the Coulomb energy $\langle H\rangle _{0}\equiv \langle \Phi
_{0}|H|\Phi _{0}\rangle $ with (\ref{ProjeHamilA}). Most terms are SU(2)
invariant, but there exists a noninvariant term,%
\begin{equation}
\langle H_{\text{noninv}}\rangle _{0}=2\pi |uv|^{2}V_{0;\text{X}}^{++++}.
\label{NoninTerm}
\end{equation}%
Hence the SU(2) symmetry is broken explicitly into the U(1) symmetry by
Coulomb interactions.

The energy is minimized either $u=0$ or $v=0$ due to the term (\ref%
{NoninTerm}), corresponding to the state%
\begin{equation}
|\Phi _{0}^{\downarrow }\rangle =\prod\limits_{n}c_{-}^{\downarrow \dagger
}\left( n\right) \left\vert 0\right\rangle \quad \text{or\quad }|\Phi
_{0}^{\uparrow }\rangle =\prod\limits_{n}c_{+}^{\uparrow \dagger }\left(
n\right) \left\vert 0\right\rangle .
\end{equation}%
The ground state is either $|\Phi _{0}^{\downarrow }\rangle $ or $|\Phi
_{0}^{\uparrow }\rangle $, though there exists still the Z$_{2}$ symmetry.
The energy barrier is of the order of the Coulomb energy, which is much
larger than the thermal energy.

We conclude as follows: The two levels split explicitly by an extrinsic
Zeeman effect if exits. Then the first energy level is up-spin polarized,
and the second energy level is down-spin polarized. Even without such an
extrinsic Zeeman effect, driven by the Coulomb exchange interaction, the
spontaneous breakdown of the Z$_{2}$ symmetry turns the system into a QH
ferromagnet\cite{BookDasSarma,BookEzawa}. It is reasonable to call it the
Ising QH ferromagnet due to the Z$_{2}$ symmetry. In any case, a plateau
emerges at $\nu =1$, where the activation energy is of the order of the
typical Coulomb energy as in the conventional QHE\cite%
{BookDasSarma,BookEzawa}.

\subsection{$N$th Energy Level}

We next study the $N$th energy level with $N\geq 1$. It contains four
degenerate states with the SU(4) symmetry in noninteracting theory. However,
the projected density is invariant only under U(1)$\times $U(1)$\times $Z$%
_{2}$. Hence we take a set of trial functions by requiring this symmetry, 
\begin{subequations}
\label{PpinPolarState}
\begin{align}
|\Phi _{N}^{\uparrow }\rangle & =\prod\limits_{n}\left( \sin \alpha
_{N}e^{i\theta ^{\uparrow }}c_{+}^{\uparrow \dagger }\left( n\right) +\cos
\alpha _{N}e^{-i\theta ^{\uparrow }}c_{-}^{\uparrow \dagger }\left( n\right)
\right) \left\vert 0\right\rangle , \\
|\Phi _{N}^{\downarrow }\rangle & =\prod\limits_{n}\left( \cos \alpha
_{N}e^{-i\theta ^{\downarrow }}c_{+}^{\downarrow \dagger }\left( n\right)
+\sin \alpha _{N}e^{i\theta ^{\downarrow }}c_{-}^{\downarrow \dagger }\left(
n\right) \right) \left\vert 0\right\rangle ,
\end{align}%
where the phase factors $e^{i\theta ^{\uparrow }}$ and $e^{i\theta
^{\downarrow }}$ assure the U(1)$\times $U(1) symmetry. These two states are
degenerate, $\langle \Phi _{N}^{\uparrow }|H_{N}|\Phi _{N}^{\uparrow
}\rangle =\langle \Phi _{N}^{\downarrow }|H_{N}|\Phi _{N}^{\downarrow
}\rangle $, due to the Z$_{2}$ symmetry.

It is easy to determine the angle $\alpha _{N}$ by minimizing the Coulomb
energy $\langle \Phi _{N}^{\sigma }|H_{N}|\Phi _{N}^{\sigma }\rangle $.
After some calculations we find
\end{subequations}
\begin{equation}
\langle \Phi _{N}^{\uparrow }|H_{N}|\Phi _{N}^{\uparrow }\rangle =A|\sin
\alpha _{N}|^{4}+B|\cos \alpha _{N}|^{4}+C|\sin \alpha _{N}\cos \alpha
_{N}|^{2},
\end{equation}%
where
\begin{subequations}
\begin{align}
A=& V_{N;\text{D}}^{++++}-V_{N;\text{X}}^{++++},\qquad B=V_{N;\text{D}%
}^{----}-V_{N;\text{X}}^{----}, \\
C=& V_{N;\text{D}}^{++--}+V_{N;\text{D}}^{--++}-V_{N;\text{X}}^{+--+}-V_{N;%
\text{X}}^{-++-}
\end{align}%
with (\ref{G-VDX}). Here, $V_{N;\text{D}}^{++++}=V_{N;\text{D}}^{----}=V_{N;%
\text{D}}^{++--}=V_{N;\text{D}}^{--++}$ since $F_{\tau \tau ^{\prime
}}^{\sigma }(0)=1$. Furthermore, because of (\ref{G-FactoK}), $V_{N;\text{X}%
}^{+--+}$ and $V_{N;\text{X}}^{-++-}$ are exponentially smaller in $(a/\ell
_{B})^{2}$ than $V_{N;\text{X}}^{++++}$ or $V_{N;\text{X}}^{----}$, and can
be neglected. Hence
\end{subequations}
\begin{equation}
\tan ^{2}\alpha _{N}=\frac{B-C/2}{A-C/2}\simeq \frac{V_{N;\text{X}}^{----}}{%
V_{N;\text{X}}^{++++}}.
\end{equation}%
Now it follows from (\ref{G-FormFactor}) that $V_{N;\text{X}}^{----}=V_{N-1;%
\text{X}}^{++++}$. We can also see [Fig.\ref{FigProjV}(a)]%
\begin{equation}
V_{N-1;\text{X}}^{++++}>V_{N;\text{X}}^{++++},
\end{equation}%
implying that the Coulomb energy of an electron in higher Landau level is
lower. It follows that 
\begin{equation}
\tan ^{2}\alpha _{N}\simeq \frac{V_{N-1;\text{X}}^{++++}}{V_{N;\text{X}%
}^{++++}}>1.  \label{Tan}
\end{equation}%
We have depicted $\tan ^{2}\alpha _{N}$ as a function of $N$ in Fig.\ref%
{FigProjV}(b).

We conclude that, if we take the state $|\Phi _{N}^{\uparrow }\rangle $,
more electrons are present in the Dirac valley at the K point than at the K'
point since $|\sin \alpha _{N}|>|\cos \alpha _{N}|$. Namely, the valley
polarization has occurred both in $|\Phi _{N}^{\uparrow }\rangle $ and $%
|\Phi _{N}^{\downarrow }\rangle $. This can be understood physically as
follows [Fig.\ref{FigGraphLevel}(a)]: Up-spin electrons in the K point
belong to the ($N$-$1$)th Landau level but those in the K' point belong to
the $N$th Landau level. It is easier to fill Landau sites at the K point
because the Coulomb energy is lower in higher Landau levels. The valley
polarization disappears as $N\rightarrow \infty $, since $V_{N+1;\text{X}%
}^{----}=V_{N;\text{X}}^{----}$ in the limit [Fig.\ref{FigProjV}(b)].

An extrinsic Zeeman effect open a gap between these two spin polarized
states. Even without such an effect, driven by the Coulomb exchange
interaction, the spontaneous breakdown of the Z$_{2}$ symmetry turns the
system into a QH ferromagnet. Note that it has still the 2-fold degeneracy.
In any case, a plateau emerges at $\nu =4N$ [Fig.\ref{FigSplit}(c)], where
the activation energy is of the order of the typical Coulomb energy as in
the conventional QHE\cite{BookDasSarma,BookEzawa}.

\section{Multilayer Graphene Systems}

\label{SecMultiGraph}

\subsection{Bilayer Graphene (Bernal Stacking)}

We proceed to generalize the above analysis to a bilayer graphene, which is
a system made of two coupled hexagonal lattices according to the Bernal
stacking. In the absence of magnetic field, the low-energy spectrum of the
bilayer graphene is known\cite{Nov3,McCann,Guinea,Koshino} to be parabolic,%
\begin{equation}
\mathcal{E}\left( k\right) \propto |\boldsymbol{k}|^{2},  \label{G-EnergBL}
\end{equation}%
near the K and K' points. In the presence of the magnetic field, we can
reformulate the model Hamiltonian\cite{McCann,Guinea,Koshino} as the
generalized Dirac Hamiltonian defined by%
\begin{equation}
H^{\pm }=\text{diag.}\left( \sqrt{Q_{\pm }Q_{\pm }},-\sqrt{Q_{\pm }Q_{\pm }}%
\right) ,
\end{equation}%
together with%
\begin{equation}
Q_{+}=\left( 
\begin{array}{cc}
0 & A^{\dagger } \\ 
A & 0%
\end{array}%
\right) ,\quad Q_{-}=\left( 
\begin{array}{cc}
0 & A \\ 
A^{\dagger } & 0%
\end{array}%
\right) .
\end{equation}%
Here, $A=\hbar \omega _{c}a^{2}$, with $a$ given by (\ref{G-OperaA}). We
also consider the generalized Pauli Hamiltonian
\begin{subequations}
\begin{align}
H_{\text{P}}^{+}\equiv & Q_{+}Q_{+}=(\hbar \omega _{c})^{2}\left( 
\begin{array}{cc}
a^{2\dagger }a^{2} & 0 \\ 
0 & a^{2}a^{2\dagger }%
\end{array}%
\right) , \\
H_{\text{P}}^{-}\equiv & Q_{-}Q_{-}=(\hbar \omega _{c})^{2}\left( 
\begin{array}{cc}
a^{2}a^{2\dagger } & 0 \\ 
0 & a^{2\dagger }a^{2}%
\end{array}%
\right) .
\end{align}%
We may switch off the magnetic field in this formula, and reproduce the
energy spectrum (\ref{G-EnergBL}).

The eigenvalue of the Hamiltonian $H^{\pm }$ is derived as
\end{subequations}
\begin{equation}
H^{\pm }|N\rangle =\left( \mathcal{E}_{N}^{\pm \uparrow },\mathcal{E}%
_{N}^{\pm \downarrow },-\mathcal{E}_{N}^{\pm \uparrow },-\mathcal{E}%
_{N}^{\pm \downarrow }\right) |N\rangle  \label{G-EnergEigen}
\end{equation}%
with the eigenstate $|N\rangle =(N!)^{-1/2}(a^{\dagger })^{N}|0\rangle $,
where
\begin{subequations}
\begin{align}
\mathcal{E}_{N}^{+\uparrow }& =\mathcal{E}_{N}^{-\downarrow }=\hbar \omega
_{c}\sqrt{N\left( N-1\right) }, \\
\mathcal{E}_{N}^{+\downarrow }& =\mathcal{E}_{N}^{-\uparrow }=\hbar \omega
_{c}\sqrt{\left( N+2\right) \left( N+1\right) }
\end{align}%
for $N=0,1,2,3$, as illustrated in Fig.\ref{FigGraphBL}. It is interesting
that two Landau levels mix to create one nonzero-energy level, but that four
Landau levels mix to create the zero-energy level. Thus there exists the
4-fold degeneracy in the nonzero-energy level but the 8-fold degeneracy in
the zero-energy state, as results in the bold-face series (\ref{G-SerieBL}).
This agrees with the previous result\cite{Nov3,McCann,Guinea}.

\begin{figure}[t]
\begin{center}
\includegraphics[width=0.9\textwidth]{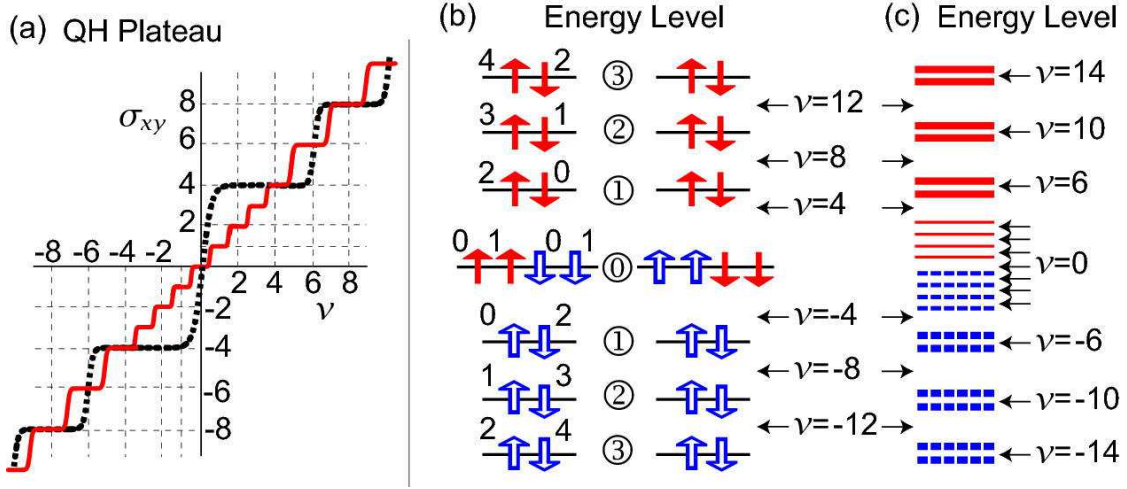}
\end{center}
\caption{\baselineskip=12pt{}(Color online) (a) The QH conductivity in bilayer graphene. The
dotted black curve shows the sequence $\protect\nu =\pm 4,\pm 8,\pm
12,\cdots $, while the solid red curve the sequence $\protect\nu =0,\pm
1,\pm 2,\pm 3,\pm 4,\pm 6,\pm 8,\cdots $. (b) The energy level within
noninteracting theory. The spin is indicated by a solid red (open blue)
arrow for electron (hole) at the K and K' points. The number attached to a
solid red (open blue) arrow shows from which Landau level the electron
(hole) comes from. (c) When Coulomb interactions are included, the
zero-energy level splits into eight nondegenerate subbands, while each
nonzero-energy level splits into two 2-fold degenerate subbands with the
exact U(1) symmetry.}
\label{FigGraphBL}
\end{figure}

We include Coulomb interactions. The $N$th energy level ($N\neq 0$) has the
same structure as in the monolayer graphene system. Coulomb interactions
make each energy level split into two subbands.

We discuss the zero-energy state in some details. There are electron-hole
pairs coming from the $N=0$ Landau level and the $N=1$ Landau level for
electrons and holes [Fig.\ref{FigGraphBL}(b)]. At the K point there are four
gap functions $\Delta _{00}\left( \boldsymbol{k}\right) $, $\Delta _{11}\left( 
\boldsymbol{k}\right) $, $\Delta _{01}\left( \boldsymbol{k}\right) $ and $\Delta
_{10}\left( \boldsymbol{k}\right) $, 
\end{subequations}
\begin{equation}
\Delta _{NN^{\prime }}\left( \boldsymbol{k}\right) =\int {\!}d^{2}k^{\prime
}\,V^{\text{eff}}(\boldsymbol{k}^{\prime }-\boldsymbol{k)}\langle \psi _{\text{e}%
}^{N\uparrow \dag }(\boldsymbol{k}^{\prime })\psi _{\text{h}}^{N^{\prime
}\downarrow \dag }(\boldsymbol{k}^{\prime })\rangle ,
\end{equation}%
where $\psi _{\text{e}}^{N\uparrow \dag }$ and $\psi _{\text{h}}^{N^{\prime
}\uparrow \dag }$ are creation operators of electrons in the $N$th Landau
level and holes in the $N^{\prime }$th Landau level, respectively. Repeating
similar analysis we have made in Section \ref{SecExciton}, we solve the gap
equations at $T=0$ as
\begin{subequations}
\begin{align}
\Delta _{00}\left( \boldsymbol{k}\right) =& \Delta _{0}, \\
\Delta _{11}\left( \boldsymbol{k}\right) =& \frac{3}{4}\Delta _{0}, \\
\Delta _{01}\left( \boldsymbol{k}\right) =& \Delta _{10}\left( \boldsymbol{k}\right)
=\frac{1}{2}\Delta _{0},
\end{align}%
where $\Delta _{0}$ is given by (\ref{GapAtZero}); $\Delta _{0}=\pi \sqrt{%
\pi /2}(e^{2}/4\pi \varepsilon \ell _{B})$. Thus, the 8-fold degenerate
zero-energy level splits into four 2-fold degenerated subbands, producing
plateaux at $\nu =\pm 2,\pm 4$. Furthermore we have Ising QH ferromagnets at 
$\nu =\pm 1,\pm 3$. Consequently, we predict the full series (\ref{G-SerieBL}%
).

\subsection{Trilayer Graphene (Rhombohedral Stacking)}

The above analysis is applicable also to a trilayer graphene with the ABC
stacking (rhombohedral stacking). In the absence of the magnetic field, the
low-energy spectrum of trilayer graphene has been argued\cite{Guinea} to be
cubic, 
\end{subequations}
\begin{equation}
\mathcal{E}\left( k\right) \propto |\boldsymbol{k}|^{3},  \label{G-EnergTL}
\end{equation}%
near the K and K' points. In the presence of the magnetic field, we can
reformulate the model Hamiltonian\cite{Guinea} as the generalized Dirac
Hamiltonian as%
\begin{equation}
H^{\pm }=\text{diag.}\left( \sqrt{Q_{\pm }Q_{\pm }},-\sqrt{Q_{\pm }Q_{\pm }}%
\right) ,
\end{equation}%
together with%
\begin{equation}
Q_{+}=\left( 
\begin{array}{cc}
0 & A^{\dagger } \\ 
A & 0%
\end{array}%
\right) ,\quad Q_{-}=\left( 
\begin{array}{cc}
0 & A \\ 
A^{\dagger } & 0%
\end{array}%
\right) .
\end{equation}%
Here, $A=\hbar \omega _{c}a^{3}$, where $a$ given by (\ref{G-OperaA}).

The eigenvalue of the Hamiltonian $H^{\pm }$ is derived as in (\ref%
{G-EnergEigen}) with
\begin{subequations}
\begin{align}
\mathcal{E}_{N}^{+\uparrow }& =\mathcal{E}_{N}^{-\downarrow }=\hbar \omega
_{c}\sqrt{N\left( N-1\right) (N-2)}, \\
\mathcal{E}_{N}^{+\downarrow }& =\mathcal{E}_{N}^{-\uparrow }=\hbar \omega
_{c}\sqrt{\left( N+3\right) \left( N+2\right) \left( N+1\right) }
\end{align}%
for $N=0,1,2,3$, as illustrated in Fig.\ref{FigGraphBL}. There exists the
4-fold degeneracy in the nonzero-energy level but the 12-fold degeneracy in
the zero-energy state, as results in the bold-face series (\ref{G-SerieTL}).
This agrees with the previous result\cite{Guinea}. 
\begin{figure}[t]
\begin{center}
\includegraphics[width=0.92\textwidth]{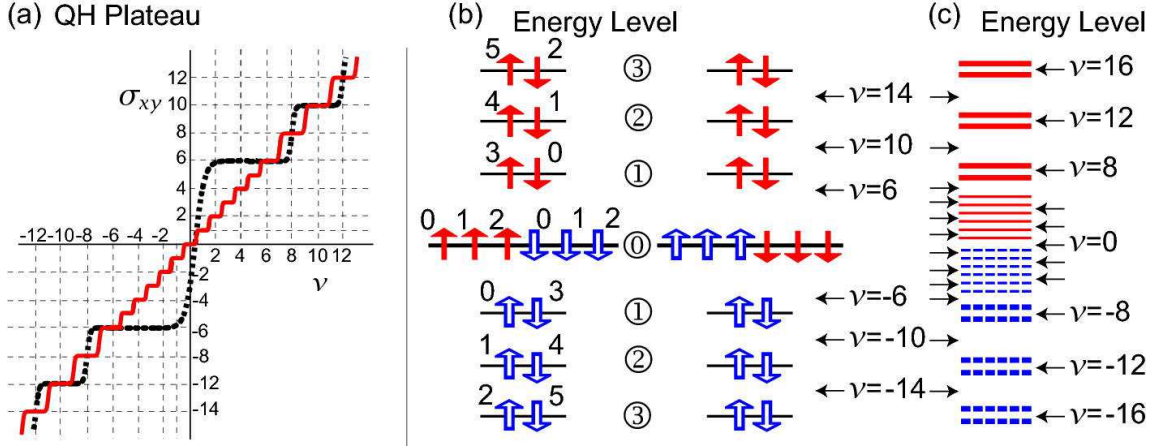}
\end{center}
\caption{\baselineskip=12pt{}(Color online) (a) The QH conductivity in trilayer graphene. The
dotted black curve shows the sequence $\protect\nu =\pm 6$, $\pm 10$, $\pm 14
$, $\cdots $, while the solid red curve the sequence $\protect\nu =0$, $\pm 1
$, $\pm 2$, $\pm 3$, $\pm 4$, $\pm 5$, $\pm 6$, $\pm 8$, $\pm 10$, $\cdots $%
. (b) The energy level within noninteracting theory. The spin is indicated
by a solid red (open blue) arrow for electron (hole) at the K and K' points.
The number attached to a solid red (open blue) arrow shows from which Landau
level the electron (hole) comes from. (c) When Coulomb interactions are
included, the zero-energy level splits into twelve nondegenerate subbands,
while each nonzero-energy level splits into two 2-fold degenerate subbands
with the exact U(1) symmetry.}
\label{FigGraphTL}
\end{figure}

We briefly argue how the energy\ spectrum is modified by Coulomb
interactions. The effective Coulomb potential depends on the spin and the
valley degree of freedom precisely by the same mechanism. The $N$th energy
level ($N\neq 0$) has the same structure as in the monolayer graphene
system. Hence, each energy level splits into two subbands. The ground state
is describe by a formula similar to (\ref{PpinPolarState}), where the valley
polarization is occurred. There are electron-hole pairs in the zero-energy
state, which make excitonic condensation by the same mechanism as in the
monolayer and bilayer cases. Thus, the 12-fold degenerate zero-energy level
splits into six 2-fold degenerated subbands, producing plateaux at $\nu =\pm
2,\pm 4,\pm 6$. Furthermore we have Ising QH ferromagnets at $\nu =\pm 1,\pm
3,\pm 5$. Consequently, we predict the full series (\ref{G-SerieTL}).

\section{Discussions}

\label{SecDiscu}

The most intriguing property of the graphene system is that the intrinsic
Zeeman energy is precisely one half of the cyclotron energy for electrons
and holes. It leads to the symmetry group SU(4) in the noninteracting
theory, where the Hall plateau emerges at $\nu =\pm 2,\pm 6,\pm 10,\cdots $.
This series is the first experimental result\cite{Nov2} of the QHE in
graphene. When Coulomb interactions are included, the symmetry SU(4) is
broken so that the Hall plateau emerges at $\nu =0,\pm 1,\pm 4,\pm 8,\cdots $%
. This series has been found experimentally\cite{Zhang06L} when larger
magnetic field is applied. We have shown that the Ising QH ferromagnets
appear at $\nu =\pm 1$ due to a BCS-type condensation of electron-hole pairs.

We have emphasized that one energy level contains up-spin and down-spin
electrons coming from two neighboring Landau levels. Since Coulomb
interactions are different for electrons in different Landau levels, we have
derived a remarkable consequence that the effective Coulomb potential
depends on the spin and the valley degree of freedom [Figs.\ref%
{FigProjeCoulo} and \ref{FigProjN}]. As a result, the valley polarization is
occurred on the ground state. We wish to explore new physics associated with
this peculiar Coulomb interaction in forthcoming papers.

\section*{Acknowledgement}

The work was in part supported by Grants-in-Aid for Scientific Research from
Ministry of Education, Science, Sports and Culture (Nos.070500000466).

\appendix

\section{Dirac Electrons in Graphene}

\label{SecGraphDirac}

In this appendix, we derive the second-quantized Dirac Hamiltonian (\ref%
{GraphHamilD}) together with (\ref{DiracHamilD}) for electrons in graphene.
We start with a study on the quantum-mechanical states of electrons based on
the $\boldsymbol{k}\cdot \boldsymbol{p}$ approximation\cite{Slonczewski,Ajiki}.

First of all, the wave functions $f_{S}^{\tau }(\boldsymbol{x})$ are given in
terms of envelope functions $F_{S}^{\tau }(\boldsymbol{x})$ as\cite{AndoReview}
\end{subequations}
\begin{equation}
f_{S}^{\text{K}}(\boldsymbol{x})=e^{i\boldsymbol{K}\cdot \boldsymbol{x}}F_{S}^{\text{K}}(%
\boldsymbol{x}),\qquad f_{S}^{\text{K'}}(\boldsymbol{x})=e^{-i\boldsymbol{K}\cdot 
\boldsymbol{x}}F_{S}^{\text{K'}}(\boldsymbol{x}),  \label{WaveEnvel}
\end{equation}%
where $S$ is the site index ($S=$A$,$B) and $\tau $ is the valley index ($%
\tau =$K,K'). The envelope functions are normalized as $|F_{S}^{\text{K}}(%
\boldsymbol{x})|^{2}+|F_{S}^{\text{K'}}(\boldsymbol{x})|^{2}=1/2$ for $S=$A$,$B.
They satisfy the Schr\"{o}dinger equation\cite{AndoReview} 
\begin{equation}
H_{0}\boldsymbol{F}(\boldsymbol{x})=\mathcal{E}\boldsymbol{F}(\boldsymbol{x}),
\end{equation}%
where $\mathcal{E}$ is the eigenvalue, and the Hamiltonian $H_{0}$ is the $%
4\times 4$ matrix operator,%
\begin{equation}
H_{0}=v_{\text{F}}\left( 
\begin{array}{cccc}
0 & p_{x}-ip_{y} & 0 & 0 \\ 
p_{x}+ip_{y} & 0 & 0 & 0 \\ 
0 & 0 & 0 & p_{x}+ip_{y} \\ 
0 & 0 & p_{x}-ip_{y} & 0%
\end{array}%
\right) ,  \label{AndoHamil}
\end{equation}%
with $p_{k}=-i\hbar \partial _{k}$. Since this is block diagonal, it is
convenient to set 
\begin{subequations}
\label{KP1}
\begin{eqnarray}
H_{\text{R}}^{\text{K}} &=&v_{\text{F}}(\boldsymbol{\sigma \cdot p})=v_{\text{F}%
}\left( 
\begin{array}{cc}
0 & p_{x}-ip_{y} \\ 
p_{x}+ip_{y} & 0%
\end{array}%
\right) , \\
H_{\text{R}}^{\text{K'}} &=&v_{\text{F}}(\boldsymbol{\sigma \cdot p}^{\prime
})=\left( 
\begin{array}{cc}
0 & p_{x}+ip_{y} \\ 
p_{x}-ip_{y} & 0%
\end{array}%
\right) ,
\end{eqnarray}%
with $\boldsymbol{p}^{\prime }=(p_{x},-p_{y})$. Note that $H_{\text{R}}^{\text{K'%
}}=\sigma _{x}H_{\text{R}}^{\text{K}}\sigma _{x}$, where $\sigma _{x} $ is
the generator of the mirror symmetry. The eigenfunctions of $H_{\text{R}%
}^{\tau }$ ($\tau =$K,K') are given by\cite{AndoReview}
\end{subequations}
\begin{equation}
\boldsymbol{F}_{\sigma }^{\tau \text{;R}}(\boldsymbol{x})=\left( 
\begin{array}{c}
F_{\text{A}}^{\tau }(\boldsymbol{x}) \\ 
\sigma F_{\text{B}}^{\tau }(\boldsymbol{x})%
\end{array}%
\right) ,  \label{KP2}
\end{equation}%
with the eigenvalue $\mathcal{E}(\boldsymbol{k}^{\prime })=\sigma \hbar v_{\text{%
F}}|\boldsymbol{k}^{\prime }|$, where $\sigma $ stands for the helicity, $\sigma
=\pm $. Note that $\hbar \boldsymbol{k}^{\prime }$ is the momentum of electrons, 
$|\boldsymbol{k}^{\prime }|\ll |\boldsymbol{K}|$ measured from the K or K' point.

Let us explain the notations we have used in (\ref{KP1}) and (\ref{KP2}). We
recall that the $\boldsymbol{\sigma \cdot p}$ is the helicity operator except
for the positive normalization factor. Thus, $\boldsymbol{F}_{+}^{\text{K;R}}(%
\boldsymbol{x})$ has the positive helicity and a positive energy, while $\boldsymbol{%
F}_{-}^{\text{K;R}}(\boldsymbol{x})$ has the negative helicity and a negative
energy. Since the energy and the helicity have the same sign for the
envelope function $\boldsymbol{F}_{\sigma }^{\text{K;R}}(\boldsymbol{x})$, it
describes the right-handed Weyl fermion by definition. Hence we have put the
index "R". Similarly we have assigned the right-handed Weyl fermion at the
K' point.

There are only four independent two-component envelope functions given by (%
\ref{KP2}) with $\tau =$K,K' and $\sigma =\pm $. However, there are more
quantum-mechanical states for electrons. They are the chiral symmetric
copies; the left-handed Weyl fermions $\boldsymbol{F}_{\sigma }^{\text{K;L}}(%
\boldsymbol{x})$ at the K point and $\boldsymbol{F}_{\sigma }^{\text{K';L}}(\boldsymbol{x%
})$ at the K' point. The corresponding Hamiltonians and envelope functions
are constructed by the chiral transformation generated by the Pauli matrix $%
\sigma _{z}$ as $H_{\text{L}}^{\text{K}}=\sigma _{z}H_{\text{R}}^{\text{K}%
}\sigma _{z}=-v_{\text{F}}(\boldsymbol{\sigma \cdot p})$, $H_{\text{L}}^{\text{K'%
}}=\sigma _{z}H_{\text{R}}^{\text{K'}}\sigma _{z}=-v_{\text{F}}(\boldsymbol{%
\sigma \cdot p}^{\prime })$, and%
\begin{equation}
\boldsymbol{F}_{\sigma }^{\tau \text{;L}}(\boldsymbol{x})=\sigma _{z}\boldsymbol{F}%
_{\sigma }^{\tau \text{;R}}(\boldsymbol{x})=\left( 
\begin{array}{c}
F_{\text{A}}^{\tau }(\boldsymbol{x}) \\ 
-\sigma F_{\text{B}}^{\tau }(\boldsymbol{x})%
\end{array}%
\right) .
\end{equation}%
Note the the energy and the helicity have the opposite sign for the
left-handed Weyl fermion.

The energy spectrum is symmetric between the positive and negative energy
states. There exists one electron per one carbon and the band-filling factor
is 1/2 in graphene. Namely, all negative-energy states are filled up, as is
a reminiscence of the Dirac sea. Hence, we have electrons and holes as
physical excitations.

In this way there are eight types of quantum-mechanical states for electrons
in graphene, corresponding to the spin degree of freedom ($\sigma =\pm $),
the electron-hole degree of freedom and the valley degree of freedom ($\tau
= $K,K'). To carry out the second quantization it is necessary to use all
these quantum-mechanical states together with the four Hamiltonians. We
arrange them as $H_{\text{D}}^{\text{K}}=v_{\text{F}}\boldsymbol{\sigma \cdot p}%
\gamma _{5}$, $H_{\text{D}}^{\text{K'}}=-v_{\text{F}}\boldsymbol{\sigma \cdot p}%
^{\prime }\gamma _{5}$ where we have introduced the Dirac $\gamma _{5}$
matrix in the Weyl representation,%
\begin{equation}
\gamma _{5}=\left( 
\begin{array}{cc}
1 & 0 \\ 
0 & -1%
\end{array}%
\right) .
\end{equation}%
They are summarized into the Dirac Hamiltonian (\ref{DiracHamilD}) in text,
or 
\begin{equation}
H_{\text{D}}^{\tau }=v_{\text{F}}\left( \tau \sigma _{x}p_{x}+\sigma
_{y}p_{y}\right) \gamma _{5},
\end{equation}%
where $\tau =+$ ($-$) for the K (K') point.

According to the standard prescription the second-quantized Hamiltonian is
constructed as in (\ref{GraphHamilD}), or $H=\sum_{\tau =\pm }\int
\!d^{2}x\,\Psi _{\tau }^{\dagger }(\boldsymbol{x})H_{\text{D}}^{\tau }\Psi
_{\tau }(\boldsymbol{x})$, where $H_{\text{D}}^{\tau }$ is the quantum
mechanical Hamiltonian. The field operator is expanded as $\Psi _{\tau }(%
\boldsymbol{x})=\Psi _{\text{e}\tau }(\boldsymbol{x})+\Psi _{\text{h}\tau }(\boldsymbol{x%
})$, with
\begin{subequations}
\begin{align}
\Psi _{\text{e}\tau }(\boldsymbol{x})& =\sum_{\sigma =\pm }\int {\frac{%
d^{2}k^{\prime }}{2\pi }}c_{\tau }^{\sigma }(\boldsymbol{k}^{\prime })u_{\sigma
}^{\tau }(\boldsymbol{k}^{\prime })e^{i\boldsymbol{k^{\prime }x}}, \\
\Psi _{\text{h}\tau }(\boldsymbol{x})& =\sum_{\sigma =\pm }\int {\frac{%
d^{2}k^{\prime }}{2\pi }}d_{\tau }^{\sigma \dag }(\boldsymbol{k}^{\prime
})v_{\sigma }^{\tau }(\boldsymbol{k}^{\prime })e^{-i\boldsymbol{k^{\prime }x}}.
\end{align}%
We have introduced the annihilation operator $c_{\tau }^{\sigma }(\boldsymbol{k}%
^{\prime })$ of an electron with the eigenfunction $u_{\sigma }^{\tau }(%
\boldsymbol{k}^{\prime })$, and the creation operator $d_{\tau }^{\sigma \dag }(%
\boldsymbol{k}^{\prime })$ of a hole with the eigenfunction $v_{\sigma }^{\tau }(%
\boldsymbol{k}^{\prime })$, where we have set $u_{+}^{\tau }(\boldsymbol{k}^{\prime
})=\boldsymbol{F}_{+}^{\tau \text{;R}}(\boldsymbol{k}^{\prime })$, $u_{-}^{\tau }(%
\boldsymbol{k}^{\prime })=\boldsymbol{F}_{-}^{\tau \text{;L}}(\boldsymbol{k}^{\prime })$%
, $v_{+}^{\tau }(\boldsymbol{k}^{\prime })=\boldsymbol{F}_{+}^{\tau \text{;L}}(-%
\boldsymbol{k}^{\prime })$ and $v_{-}^{\tau }(\boldsymbol{k}^{\prime })=\boldsymbol{F}%
_{-}^{\tau \text{;R}}(-\boldsymbol{k}^{\prime })$ in accord with the standard
notation in the Dirac theory. In passing, the field operator $\psi _{\tau }(%
\boldsymbol{x})$ for electrons in graphene is to be constructed with the use of
the wavefuncion (\ref{WaveEnvel}), and hence is given by (\ref{RelatWE}) in
text.

\section{Landau-Level Projection}

\label{AppenLLProje}

In this appendix we derive the formula (\ref{DensiInN})\ for the projected
density operator $\rho _{N}(\boldsymbol{q})$. We first review how the
Landau-level projection is made in the conventional QHE, where the density
operator is given by $\rho (\boldsymbol{x})=\psi ^{\dagger }(\boldsymbol{x})\psi (%
\boldsymbol{x}).$ We consider electrons confined to the $N$th Landau level,
where Fock states are given by (\ref{G-FockNn}). The field operator is
expanded as
\end{subequations}
\begin{equation}
\psi (\boldsymbol{x})=\sum_{n}\langle \boldsymbol{x}|N,n\rangle c(n),
\label{FieldExpanB}
\end{equation}%
with $c(n)$ the annihilation operator of electrons acting on the state $%
|N,n\rangle $, $\{c(n),c^{\dagger }(m)\}=\delta _{nm}$. The projected
density operator is%
\begin{equation}
\rho _{N}(\boldsymbol{x})=\psi ^{\dagger }(\boldsymbol{x})\psi (\boldsymbol{x}%
)=\sum_{mn}\langle N,m|\boldsymbol{x}\rangle \langle \boldsymbol{x}|N,n\rangle
c^{\dagger }(m)c(n).
\end{equation}%
Its Fourier transformation is%
\begin{equation}
\rho _{N}(\boldsymbol{q})=\frac{1}{2\pi }\sum_{mn}\int \!d^{2}x\,e^{-i\boldsymbol{q}%
\boldsymbol{x}}\langle N,m|\boldsymbol{x}\rangle \langle \boldsymbol{x}|N,n\rangle
c^{\dagger }(m)c(n).
\end{equation}%
Here, we decompose the coordinate into the guiding center and the relative
coordinate, $\boldsymbol{x}=\boldsymbol{X}+\boldsymbol{R}$, where $\boldsymbol{X}$ and $%
\boldsymbol{R}$ act on the Fock states $|n\rangle $ and $|N\rangle $,
respectively. Since they commute each other, we obtain%
\begin{equation}
\rho _{N}(\boldsymbol{q})=\frac{1}{2\pi }\sum_{mn}\langle N|e^{-i\boldsymbol{qR}%
}|N\rangle \langle m|e^{-i\boldsymbol{qX}}|n\rangle c^{\dagger }(m)c(n),
\label{ProjeDensiLanda}
\end{equation}%
or $\rho _{N}(\boldsymbol{q})=F_{N}(\boldsymbol{q})\hat{\rho}(\boldsymbol{q})$, with 
\begin{subequations}
\begin{align}
& F_{N}(\boldsymbol{q})=\langle N|e^{-i\boldsymbol{qR}}|N\rangle ,
\label{LLFormFacto} \\
& \hat{\rho}(\boldsymbol{q})=\frac{1}{2\pi }\sum_{mn}\langle m|e^{-i\boldsymbol{qX}%
}|n\rangle c^{\dagger }(m)c(n).
\end{align}%
We call $F_{N}(\boldsymbol{q})$ the Landau-level form factor, and $\hat{\rho}(%
\boldsymbol{q})$ the bare density operator.

In the graphene QHE one energy level contain four different types of
electrons described by (\ref{G-ProjeElect}). For instance, the projected
density operator for up-spin electrons is given by
\end{subequations}
\begin{align}
\rho _{N}^{\uparrow }(\boldsymbol{x})=& \langle N,m|\boldsymbol{x}\rangle \langle 
\boldsymbol{x}|N,n\rangle c_{+}^{\uparrow \dag }(m)c_{+}^{\uparrow }(n)+\langle
N-1,m|\boldsymbol{x}\rangle \langle \boldsymbol{x}|N-1,n\rangle c_{-}^{\uparrow \dag
}(m)c_{-}^{\uparrow }(n)  \notag \\
& +e^{-2i\boldsymbol{Kx}}\langle N,m|\boldsymbol{x}\rangle \langle \boldsymbol{x}%
|N-1,n\rangle c_{+}^{\uparrow \dag }(m)c_{-}^{\uparrow }(n)+e^{2i\boldsymbol{Kx}%
}\langle N-1,m|\boldsymbol{x}\rangle \langle \boldsymbol{x}|N,n\rangle
c_{-}^{\uparrow \dag }(m)c_{+}^{\uparrow }(n),  \notag
\end{align}%
where the summation over the indices $n$ and $m$ is understood. By repeating
the above process, this reads%
\begin{align}
\rho _{N}^{\uparrow }(\boldsymbol{q})=& F_{N,N}^{++}\left( \boldsymbol{q}\right) 
\hat{D}_{++}^{\uparrow \uparrow }\left( \boldsymbol{q}\right)
+F_{N-1,N-1}^{--}\left( \boldsymbol{q}\right) \hat{D}_{--}^{\uparrow \uparrow
}\left( \boldsymbol{q}\right)  \notag \\
& +F_{N,N-1}^{+-}\left( \boldsymbol{q}\right) \hat{D}_{+-}^{\uparrow \uparrow
}\left( \boldsymbol{q}\right) +F_{N-1,N}^{-+}\left( \boldsymbol{q}\right) \hat{D}%
_{-+}^{\uparrow \uparrow }\left( \boldsymbol{q}\right) ,
\end{align}%
where%
\begin{equation}
F_{N,M}^{\tau \tau ^{\prime }}\left( \boldsymbol{q}\right) =\langle N|e^{-i[%
\boldsymbol{q}+\tau \boldsymbol{K}-\tau ^{\prime }\boldsymbol{K}]\boldsymbol{R}}|M\rangle ,
\end{equation}%
and%
\begin{equation}
\hat{D}_{\tau \tau ^{\prime }}^{\sigma \sigma ^{\prime }}(\boldsymbol{q})=\frac{1%
}{2\pi }\sum_{mn}\langle m|e^{-i[\boldsymbol{q+}\tau \boldsymbol{K}-\tau ^{\prime }%
\boldsymbol{K}]\boldsymbol{X}}|n\rangle c_{\tau }^{\sigma \dag }(n)c_{\tau ^{\prime
}}^{\sigma ^{\prime }}(m).
\end{equation}%
A similarly formula is derived also for $\rho _{N}^{\downarrow }(\boldsymbol{q})$%
. Adding $\rho _{N}^{\uparrow }(\boldsymbol{q})$ and $\rho _{N}^{\downarrow }(%
\boldsymbol{q})$ we obtain (\ref{DensiInN}) in text.

\end{document}